\documentclass[
  aps,
  prb,
  twoside,
  twocolumn,
  showpacs,
  floatfix,
  superscriptaddress,
  10pt,
  preprintnumbers,
  citeautoscript,
]{revtex4-2}

\usepackage{amsmath}
\usepackage{comment}
\usepackage{glossaries}
\usepackage{graphicx}
\usepackage{here}
\usepackage[utf8]{inputenc}
\usepackage{times}
\usepackage{mathtools}
\usepackage{physics}
\usepackage{arydshln}
\usepackage{colortbl}
\usepackage{array}
\usepackage{dcolumn}
\usepackage[english]{babel}

\DeclareUnicodeCharacter{2212}{-}

\usepackage{amssymb}
\usepackage{arydshln}
\usepackage{amsmath}
\usepackage{units}
\usepackage{xcolor,colortbl}
\definecolor{LightCyan}{rgb}{0.88,1,1}
\definecolor{Gray}{gray}{0.85}
\newacronym{cbm}{CBM}{conduction band minimum}
\newacronym{dft}{DFT}{density functional theory}
\newacronym{dos}{DOS}{density of states}
\newacronym{dwf}{DWF}{Debye--Waller factor}
\newacronym{ks}{KS}{Kohn--Sham}
\newacronym{hr}{HR}{Huang--Rhys}
\newacronym{hrf}{HRF}{Huang--Rhys factor}
\newacronym{ifc}{IFC}{interatomic force constant}
\newacronym{ipr}{IPR}{inverse participation ratio}
\newacronym{pl}{PL}{photoluminescence}
\newacronym{psb}{PSB}{phonon sideband}
\newacronym{puc}{PUC}{primitive unit cell}
\newacronym{si}{SI}{supplemental information}
\newacronym{vbm}{VBM}{valence band maximum}
\newacronym{zpl}{ZPL}{zero-phonon line}
\newacronym{spe}{SPE}{single photon emitter}

\usepackage[colorlinks=true]{hyperref}
\hypersetup{
    unicode=false,          
    pdftoolbar=true,        
    pdfmenubar=true,        
    pdffitwindow=false,     
    pdfstartview={FitH},    
    pdfnewwindow=true,      
    colorlinks=true,        
    linkcolor=blue,         
    citecolor=blue,         
    filecolor=blue,         
    urlcolor=blue           
}

\usepackage{xcolor}
\usepackage{booktabs}
\AtBeginDocument{
\heavyrulewidth=.08em
\lightrulewidth=.05em
\cmidrulewidth=.03em
\belowrulesep=.65ex
\belowbottomsep=0pt
\aboverulesep=.4ex
\abovetopsep=0pt
\cmidrulesep=\doublerulesep
\cmidrulekern=.5em
\defaultaddspace=.5em
}

\graphicspath{{./figures/}}

\begin{document}

\title{Sulfur in diamond and its effect on the creation of nitrogen-vacancy defect from \textit{ab initio} simulations}
%

\newcommand{\Wigner}{HUN-REN Wigner Research Centre for Physics, P.O. Box 49, H-1525 Budapest, Hungary}
\newcommand{\BME}{Department of Atomic Physics, Institute of Physics, Budapest University of Technology and Economics, M\H{u}egyetem rakpart 3., H-1111 Budapest, Hungary}
\newcommand{\MTA}{ MTA-WFK "Lendület" Momentum Semiconductor Nanostructures Research Group, P.O. Box 49, H-1525 Budapest, Hungary}

\author{Nima Ghafari Cherati}
\affiliation{\Wigner}
\affiliation{\BME}

\author{Anton Pershin}
\affiliation{\Wigner}
\affiliation{\BME}

\author{\'Ad\'am Gali}
\affiliation{\Wigner}
\affiliation{\BME}
\affiliation{\MTA}

\begin{abstract}
The negatively charged nitrogen-vacancy (NV) center is one of the most significant and widely studied defects in diamond that plays a prominent role in quantum technologies.
The precise engineering of the location and concentration of NV centers is of great importance in quantum technology applications. To this end, irradiation techniques such as nitrogen-molecule ion implantation are applied. Recent studies have reported enhanced NV center creation and activation efficiencies introduced by nitrogen molecule ion implantation in doped diamond layers, where the maximum creation efficiency at $\sim75$\% has been achieved in sulfur-doped layers. However, the microscopic mechanisms behind these observations and the limits of the efficiencies are far from understood. In this study, we employ hybrid density functional theory calculations to compute the formation energies, charge transition levels, and the magneto-optical properties of various sulfur defects in diamond where we also consider the interaction of sulfur and hydrogen in chemical vapor-deposited diamond layers. Our results imply that the competition between the donor substitutional sulfur and the hyper-deep acceptor sulfur-vacancy complex is an important limiting factor on the creation efficiency of the NV center in diamond. However, both species are able to trap interstitial hydrogen from diamond, which favorably mediates the creation of NV centers in chemical vapor-deposited diamond layers.
\end{abstract}

\maketitle
\section{Introduction}
\label{sec:introduction}

In recent years, there has been an ever-growing interest in quantum technologies that are based on the physical realization of controllable two-level quantum mechanical systems, \textit{viz.}, quantum bits or qubits.
Defect spins in solids have emerged as a promising platform among various solutions to this end~\cite{Zhang2020}. The negatively charged nitrogen-vacancy (NV) center in diamond with electron spin $S=1$ stands out within the family of solid-state defect qubits, notable for its long spin coherence times at room temperature~\cite{Gruber1997, Kurtsiefer2000NV, DOHERTY2013, Gali2019}. Quantum technology applications demand the precise placement of the qubits in a scalable manner. Motivated by the success of n-type doping of silicon in the semiconductor industry by dopant implantation, the target defect qubits in solids can also be introduced by ion implantation, which allows for controlled defect concentrations. Specifically, nitrogen molecule ions are typically implanted into pure chemical vapor-deposited (CVD) diamond (e.g., Ref.~\onlinecite{Meijer2005}) and then annealing is applied to create NV defects by the combination of the immobile substitutional nitrogen and the mobile vacancy. The substitutional nitrogen in diamond acts as a donor and contributes an electron to the NV defect. Thus, the implanted nitrogen serves a dual role in the creation of the negatively charged NV defect in diamond: it not only facilitates the creation of the NV defect but also donates an electron to the NV defect. The creation efficiency (i.e., the ratio of successfully created negatively-charged NV defect per number of nitrogen molecule ions) of this method is $\sim1$\% (Ref.~\onlinecite{Meijer2005}). \textit{Ab initio} calculations showed~\cite{Deak2014} that the formation of a divacancy, involving two vacancies, is more probable than the combination of the substitutional nitrogen (N$_\text{s}$) and a vacancy (V). As a result,  vacancy aggregates are more likely to form than NV defects, which could explain the low creation efficiency of NV centers with this method.

Recently, a major breakthrough has been made in the creation efficiency of NV centers by doping diamond with specific elements prior to nitrogen molecule ion implantation~\cite{Luhmann2019}. In particular, pre-doping CVD diamond with phosphorus, oxygen, and sulfur has resulted in creation yields of 57.9$\%$, 69.3$\%$, and 75.3$\%$, respectively. Among those, sulfur doped diamond not only yields the highest creation efficiency of the NV centers but also provides the longest spin coherence times. The remarkable enhancement in NV center creation efficiency has been explained by the donor nature of the dopants, which negatively charge the vacancies, thus preventing vacancy aggregation and promoting the formation of NV centers, particularly when the N$_\text{s}$ is positively charged~\cite{Luhmann2019}.  

Indeed, substitutional phosphorus (P$_\text{s}$) is a well-established n-type dopant in diamond 
~\cite{Koizumi1997, Kato2007, Koizumi2006, Koizumi2000, Butorac2008, Gheeraert2000, Goss2004, Alfieri2018}, with its donor level located 0.6~eV below the conduction band minimum ($E_\text{c}-0.6$~eV). In contrast, the role of oxygen in diamond remains less well established. In our previous theoretical study with hybrid density functional supercell calculations, we found~\cite{Thiering2016} that substitutional oxygen (O$_\text{s}$) exhibits single and double donor levels at $E_\text{c}-3.2$~eV and $E_\text{c}-4.9$~eV, respectively, which are still to be confirmed in experiments. The same study predicted that O$_\text{s}$ is the most stable  oxygen-related defect, although oxygen forms a complex with vacancies, i.e., oxygen-vacancy (OV) defect, when mobile vacancies are available in diamond. The OV defect is amphoteric, possessing both deep donor and acceptor levels in the diamond band gap. In turn, the behavior of sulfur in diamond is even less understood experimentally. A theoretical study reported the donor level of substitutional sulfur (S$_\text{s}$) at $E_\text{c}-0.79$~eV (Ref.~\onlinecite{Wang2002}), suggesting a relatively shallow donor character. However, this study relied on semilocal density functional theory (DFT) calculations which significantly underestimate the band gap of diamond. Therefore, the position of the donor level of S$_\text{s}$ needs to be revisited. More recent DFT calculations have yielded the donor level at $E_\text{c}-1.6$~eV using a method which accurately reproduces the diamond band gap (Ref.~\onlinecite{Yu2019}). Another theoretical study, also based on semilocal DFT, considered a complex of sulfur and vacancy (SV) defect~\cite{Cheng2017}. They concluded that sulfur remains at the substitutional site within the SV defect~\cite{Cheng2017}, similar to the OV defect mentioned above~\cite{Thiering2016}. This is surprising, as the sulfur atom is relatively large and might be expected to relax to the interstitial site in the complex. Indeed, semilocal DFT calculations showed that sulfur occupies the bond-center interstitial site in the SV defect, resulting in a $D_{3d}$ symmetry configuration~\cite{Goss2005, Baker2008}. This configuration, also known as the "split-vacancy", may be seen as a divacancy, where a pair of adjacent vacancies is connected by an interstitial dopant atom. 
The split-vacancy configuration also appears for the group-IV--vacancy color centers in diamond~\cite{Thiering2018}. It can be denoted as VSV for sulfur dopant or VSiV for silicon dopant, referring to the dopant's position in the diamond lattice. However, since the SiV label has been rather spread in the literature, we also prefer to use it here for sulfur in the SV defect. Furthermore, the same semilocal DFT calculations~\cite{Baker2008} also predicted that the most stable interstitial sulfur configuration has significantly higher formation energy ($\sim7$~eV) than substitutional sulfur (S$_\text{s}$), while the SV defect is favored over the S$_\text{s}$ by $\sim4$~eV in the neutral charge state. Additionally, they attributed~\cite{Baker2008} the sulfur-related W31 electron spin resonance center~\cite{Wyk1986} to the negatively charged SV defect of electron spin $S=1/2$, based on both the calculated and observed hyperfine signals of $^{33}$S ($I=/3/2$) and $^{13}$C ($I=1/2$) nuclear spins. However, we note that the relative stability of these defects as a function of the Fermi-level cannot be accurately predicted in these calculations because of the inherent band gap error in semilocal density functionals.   
      
The limited understanding of the relationship between the donor levels of dopants and the creation efficiency of the NV center, coupled with the contradicting theoretical results on sulfur defects in diamond, motivated us to carry out a systematic and thorough investigation of sulfur defects in diamond using hybrid density functional theory calculations. We consider a variety of basic defects, including interstitial and substitutional sulfur defects as well as sulfur-vacancy complexes. Given that dopants and nitrogen were implanted into CVD diamond within tens of nanometers in depth in the key experimental study~\cite{Luhmann2019}, the diamond layers are likely to contain interstitial hydrogen atoms. Therefore, we also studied the interaction of hydrogen with the selected sulfur-related defects. The methodology used to accurately predict the defect properties is detailed in Section \ref{sec:methodology}. The results on the considered defects are reported in Section \ref{sec:results}, where we also discuss their implications on the creation efficiency of NV centers in diamond. 
Finally, we summarize our findings and conclusions in Section \ref{sec:conclusion}.


\section{Methodology}
\label{sec:methodology}

The calculations were performed using \textit{ab initio} density functional theory (DFT), implemented in Vienna \textit{Ab initio} Simulation Package (VASP)~\cite{kres1} within the projector-augmented-wave (PAW)~\cite{PAW1994} method. A plane wave basis with a cutoff energy of 450~eV was employed. 
Unlike in the previous studies on sulfur defects, which used the Perdew–Burke–Ernzerhof (PBE) functional~\cite{PerdewPBE1996} to optimize the structure and obtain the electronic structure, we employed the screened hybrid density functional of Heyd, Scuseria, and Ernzerhof (HSE), with using the proposed mixing (0.25) and screening (0.2~1/\AA ) parameters for solids, labeled as HSE06~\cite{Paier2006HSE}. This approach can accurately reproduce the experimental band gap of diamond and defect levels in the gap with a typical accuracy of 0.1~eV~\cite{Deak2010}. This could be achieved because of the piecewise linearity condition, i.e., the ground-state total energy as a function of electron number is nearly linear upon electron occupation with integer electron numbers that is fulfilled both to the band edges (see a recent study in Ref.~\onlinecite{Yang2023}) and the defect states inside the band gap~\cite{Deak2010}.

The defects were created in a $4\times4\times4$ cubic 512-atom supercell. With this supercell size, accurate results can be achieved by sampling the Brillouin zone at the $\Gamma$ point. The structural relaxation was performed until reaching the convergence thresholds for total energy and ionic forces of 10$^{-5}$~eV and 10$^{-3}$~eV/\AA, respectively. The optimized lattice constants of $a_{\rm{HSE06}}=3.544$~{\AA} were obtained for the unit cell.

The relative stability of various defects was evaluated based on the calculated formation energies. Specifically, we computed the formation energy $E^q_\text{f}$ for  defects in a charge state $q$ as follows~\cite{Northrup1996},
\begin{equation}
\label{eq:Eform}
    E^q_\text{f} = E^q_\text{tot} - E_\text{bulk} - \sum_{i} n_i \mu_i + qE_\text{Fermi} + E^q_\text{corr}\text{,}
\end{equation}
where $ E^q_\text{tot}$ and $ E_\text{bulk}$ are the total energies of the defective and pristine diamond, respectively, $\mu_i$ is the chemical potential for each atom type, $i$, while, $n_i$ is a number of added ($n_i > 0 $) or removed ($n_i < 0 $) atoms relative to the bulk system. $E_\text{Fermi}$ is the Fermi-level, referenced to the valence band maximum ($E_\text{v}$). $ E^q_\text{corr}$ is a correction term for the total energies of charged supercells~\cite{Freysoldt2018}. In our study, defective supercells may contain carbon, sulfur, and hydrogen atoms. The chemical potential of carbon was calculated from diamond reference as the total energy per carbon atom. For defining the sulfur chemical potential we used the total energy of a S$_8$ crystal per sulfur atom as a reference. Finally, the hydrogen chemical potential was calculated from a slab model of (001)-($2\times1$) reconstructed and hydrogen-terminated diamond surface as we described in our previous work~\cite{Deak2014}. 

The charge transition level of a given defect can be defined as $E_{\text{Fermi}} (q1 | q2) \equiv E^{q2}_{\text{f}}=E^{q1}_{\text{f}}$ which leads to the following equation,
\begin{equation}
E_{\text{Fermi}} (q1 | q2) = [(E^{q2}
_{\text{tot}}+E^{q2}_{\text{corr}})-(E^{q1}_{\text{tot}}+E^{q1}_{\text{corr}})]/(q1-q2)\text{,}
\end{equation}
\noindent
where $E_{\text{Fermi}}$ is given with respect to $E_{\text{v}}$. We note that the charge transition levels also yield the respective photoionization threshold energies. For instance, the electron from the occupied defect level may be promoted to the conduction band by illumination when the frequency of the photon can bridge the energy gap between the total energy of the photoionized excited state with the electron in the conduction band minimum and the total energy of the ground state. Therefore, the donor charge transition levels are referenced to $E_\text{c}$ because these values provide the photoionization threshold energies. On the other hand, the electrons from the valence band maximum or deeper valence bands may be promoted to the empty defect level in the band gap by illumination. Thus, the acceptor charge transition levels are referenced to $E_\text{v}$ by providing again the photoionization threshold energies for this process.

We also studied the defect chemistry in diamond, i.e., the feasibility of defect complexation  from isolated species, where we assume at least one of the constituents of the complex to be mobile at the given annealing temperature. By comparing the formation energies of the isolated defects (A and B) with that of the complex (AB), the binding energy is defined as
\begin{equation}
\label{eq:binding}
    E_\text{b} = E^f_{A} + E^f_{B} - E^f_{AB}\text{.}
\end{equation}
A positive value of the binding energy means that the complex formation is favorable from the thermodynamic point of view in this definition, although configurational entropy may also play a role \cite{van2004first}. It is important to note that the formation energy generally depends on the position of the Fermi-level in the fundamental band gap. At a given Fermi-level, A and B defects may exist in $q_A$ and $q_B$ charge states. If the signs of $q_A$ and $q_B$ are the same then the constituent defects repel each other, thus we assume that the complex does not form because of kinetic reasons. On the other hand, if the signs of $q_A$ and $q_B$ are the opposite then the defects will attract each other which could mediate the formation of the complex if the calculated $E_\text{b}>0$. Finally, the total charge should be conserved in the defect reaction. As a consequence, if $ q_\text{diff} \equiv q_{AB}-(q_A + q_B ) >0$ at the given Fermi-level and the complex formation is favorable then a hole ($h^+$) will be ejected. If $q_\text{diff}<0$ and the complex formation is favorable then an electron ($e^-$) will be ejected in the process.

We calculate the hyperfine tensor of the paramagnetic defects with the optimized geometry which can provide a fingerprint of the defects in electron spin resonance measurements of diamond where the electron spin may interact with the proximate nuclear spins. With considering the relativistic effects, the hyperfine tensor of the nucleus consists of the Fermi-contact term and the dipole-dipole term which is given as
\begin{equation}
\label{eq:hyper}
    A^I_{ij} = \frac{1}{2S}\int n_s(r) \gamma_I \gamma_e \hbar^2 \biggr[ \Bigl( \frac{8 \pi}{3} \delta(r) \Bigl) + \Bigl( \frac{3x_ix_j}{r^5} - \frac{\delta_{ij}}{r^3} \Bigl) \biggr] d^3r \text{,}
\end{equation}
where $n_s(r)$ is the spin density of the spin state $S$, $\gamma_I$ is the nuclear Bohr magneton of nucleus $I$, and $\gamma_e$ is the electron Bohr magneton. The Fermi-contact term is proportional to the spin density localized at the place of the nucleus which may be dominant compared to the dipole-dipole term when the spin density is strongly localized around the ions in the core of the defect. We calculate the hyperfine tensor and diagonalize it to obtain hyperfine constants, 
which can be directly compared with experimental data. The Fermi-contact and the dipole-dipole terms are simply derived from the trace of the hyperfine tensors. The ratio of the Fermi-contact and dipole-dipole terms characterizes the shape of the spin density. We note that Eq.~\eqref{eq:hyper} is modified within PAW formalism and the core spin polarization correction should be added for yielding accurate hyperfine tensors, in particular, for $^{13}$C isotopes with $I=1/2$ spins~\cite{Szasz2013}.


\section{Results and discussion}
\label{sec:results}

This section is split into two main parts. In part one, we study the sulfur defects in diamond. In part two, we discuss the impact of sulfur defects on the creation efficiency of the NV center introduced by nitrogen molecular ion implantation. We also compare the impacts of oxygen defects with those of the sulfur defects on this creation efficiency. The results of oxygen, nitrogen, and intrinsic defects required for the discussion are taken from previous studies. As these two parts are tightly intertwined, we decided to place them into the same section. 

\subsection{Overview on the considered sulfur defects}
The most elemental sulfur defects are the interstitial, S$_\text{i}$, substitutional, S$_\text{s}$ and the complex with a vacancy, SV. In the context of the main goal of our study, it is important to consider that many vacancies are generated during implantation. The combination of an additional vacancy with the SV defect may lead to the formation of a larger complex, SVV. 
Therefore, we also consider SVV defect where this label indicates removal of a nearby carbon atom with the dangling bond from the core of the split-vacancy SV defect. In the key experiments~\cite{Luhmann2019} that motivated our study, the dopant implantation and the subsequent nitrogen molecular ion implantation were carried out into CVD diamond layers to a depth of a few tens of nanometers. The use of CVD layers as starting materials for generating NV centers is very general, since ultrahigh purity diamond layers can only be prepared by the chemical vapor deposition method. This technique employs atomic hydrogen, which may diffuse into the diamond. Thus, it can be assumed that interstitial hydrogen (H$_\text{i}$) may exist in the top layers of CVD diamonds. These H$_\text{i}$s may interact with sulfur defects and other defects created by implantation. Specifically, we consider the complexes of hydrogen with S$_\text{s}$ and sulfur-vacancy defects, which we label by SH, SVH, and SVVH, respectively. We note that larger hydrogen-related complexes may also occur, e.g., silicon-vacancy-hydrogen complexes~\cite{Thiering2015, Mukherjee2023} but  considering these complexes with multiple hydrogen atoms is is beyond the scope of this study.

\begin{figure*}[bth]
  \includegraphics[scale=.92]{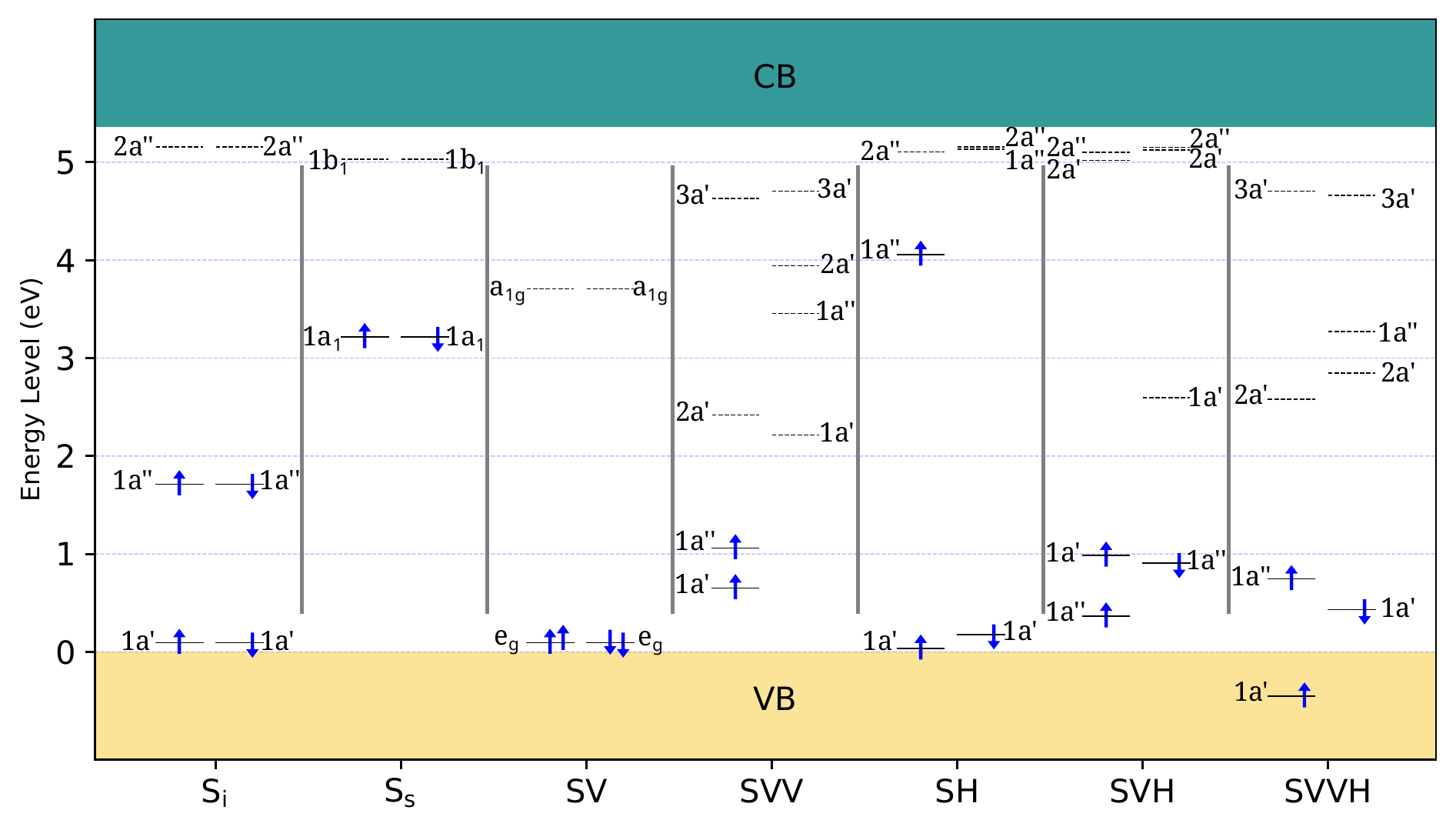}
    \caption{\label{fig:ks-states}
   Electronic structure for the ground states of the neutral defects using HSE06 functional. Kohn--Sham levels are represented by spin-up ($\uparrow$) and spin-down ($\downarrow$). The valence band (VB) and conduction band (CB) are depicted in cyan and orange, respectively. It is important to note that while the SVH structure is optimized in $C_1$ symmetry, the wavefunctions follow $C_{1h}$ symmetry. Therefore, we have labeled the states using $C_{1h}$ symmetry labels. }
\end{figure*}

\begin{figure*}[
ht!]
\includegraphics[scale=.65]{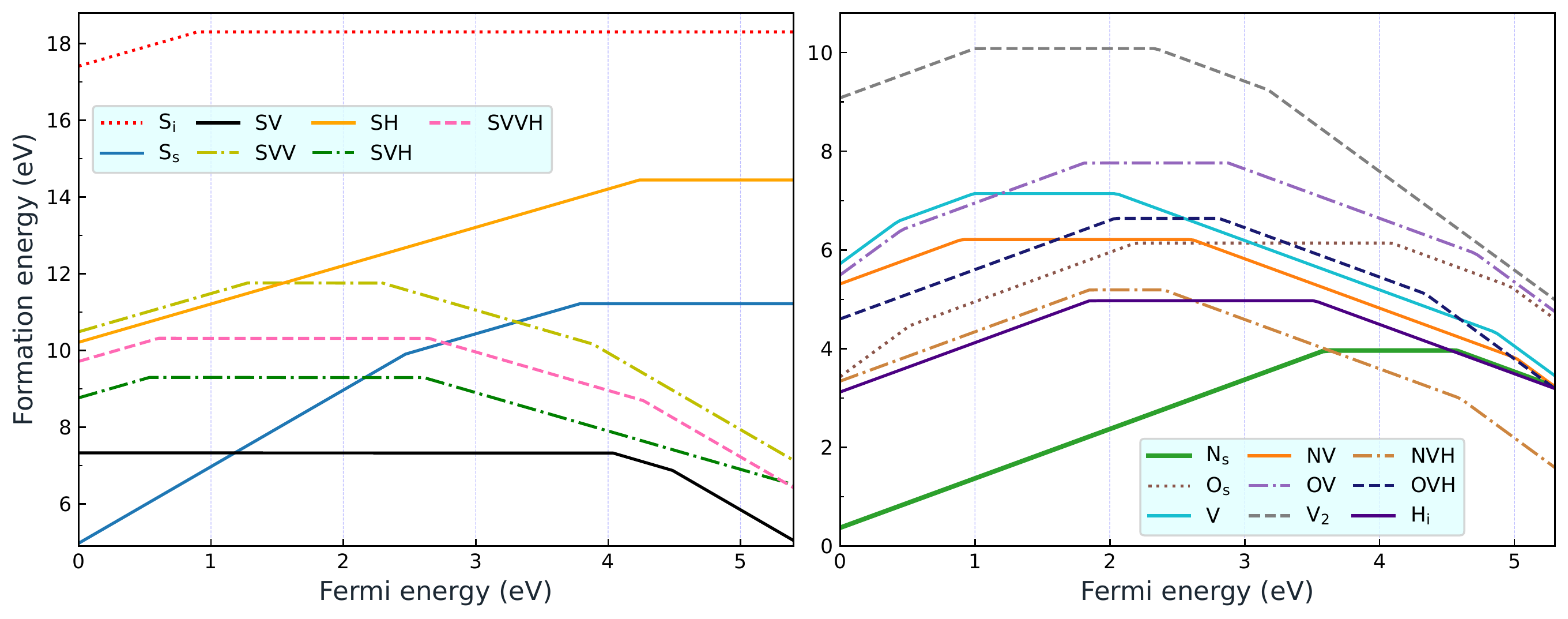}
\caption{\label{fig:formation}Formation energy of the defects as a function of the position of the Fermi-level. $E_\text{v}$ is aligned to zero for the sake of simplicity. $E_\text{c}$ is at 5.4~eV. Left panel: sulfur-related defects. Right panel: oxygen and nitrogen defects where the data are extracted from Refs.~\onlinecite{Deak2014, Thiering2016}. 
}
\end{figure*}

We first show the electronic structure of the considered neutral defect species for the most stable configurations in Fig.~\ref{fig:ks-states}. As shown, all the considered defects introduce multiple defect levels into the fundamental bandgap of diamond. We calculated the formation energies for the most stable configurations, which also revealed their stable charge states as a function of the Fermi-level. The results are plotted in the left panel of Fig.~\ref{fig:formation}. In the right panel of Fig.~\ref{fig:formation}, we also plot the formation energies of the related oxygen and nitrogen defects, which will be important in the discussion of the creation efficiency of NV centers in doped diamond.

\subsection{Interstitial sulfur defects}
\label{sec:inter}
\begin{figure*}[ht]
    \includegraphics[scale=1.12]{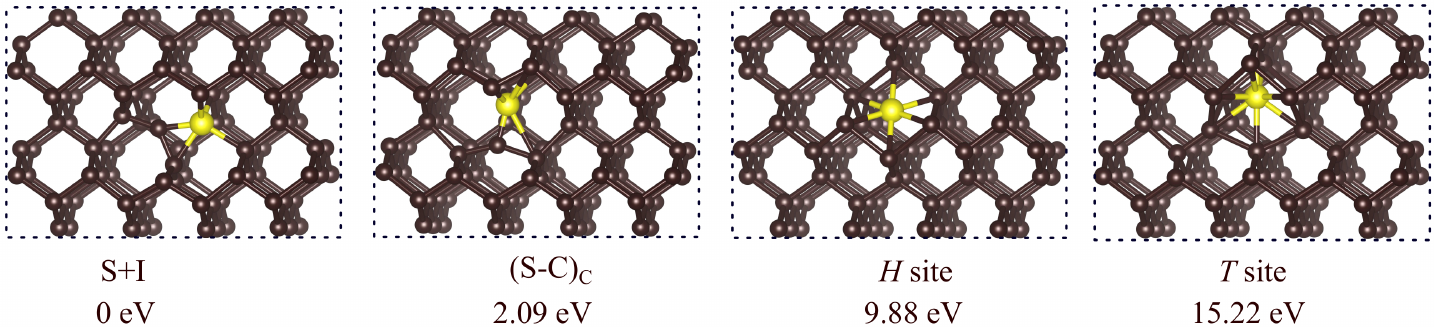}
    \caption{\label{fig:inter_structure} Geometry of optimized interstitial sulfur configurations in the order of hierarchy. (a) S+I, (b) (S-C)$_\text{C}$, (c) $H$ site, and (d) $T$ site. S and C atoms are depicted as yellow and dark spheres, respectively.}
\end{figure*}
Different sites within diamond may host sulfur as an interstitial defect. A previous DFT study has  already considered the possible configurations of sulfur interstitial ~\cite{Baker2008}, which we use as a guideline in this work. More specifically, they found that S$_\text{i}$ is least stable at the empty $T$ and $H$ sites of the diamond lattice, but is more stable in a distorted [001] split-interstitial position~\cite{Baker2008}. Furthermore, it was shown~\cite{Baker2008} that this configuration transforms into the S$_\text{s}$+C$_\text{i}$ configuration with a relatively low barrier energy, where C$_\text{i}$ is a self-interstitial carbon atom near S$_\text{s}$ in a split-interstitial position. The S$_\text{s}$+C$_\text{i}$ complex then becomes the most stable configuration of the interstitial sulfur. This type of complex formation is well-known for the boron interstitial in silicon and is often labeled as B+I complex, where 'I' denotes the self-interstitial atom (see e.g., Ref.~\onlinecite{Deak2005}). By following this syntax, the most stable interstitial sulfur may be labeled as S+I in diamond (see Fig~\ref{fig:inter_structure}(a)). We note that the relative stability of defects can depend on the choice of density functional, as the total energy depends on the defect level positions in the gap. The main reason behind this phenomenon is that the occupied defect levels with underestimated bandgap in (semi)local DFT calculations can shift significantly in hybrid DFT calculations with an enlarged bandgap. Therefore, it is useful to revisit the relative stability of interstitial sulfur configurations with hybrid DFT calculations. We found that the hierarchy of stabilities agrees well with the (semi)local DFT results, with the S+I configuration being the most stable interstitial sulfur defect. The distorted [001] split-interstitial configuration follows by 2.09~eV (see Fig.3~\ref{fig:inter_structure}(b)), while the $H$ and $T$ sites exhibit significantly higher total energies (see Fig.~\ref{fig:inter_structure}(c) and (d)).

The electronic structure of the most stable interstitial sulfur defect in the ground state is depicted in Fig.~\ref{fig:ks-states}. We find this defect to have a $C_{1h}$ ($C_s$) symmetry with a closed-shell singlet ground state and three localized states in the fundamental bandgap. The 1a$^{\prime}$ and 1a$^{\prime\prime}$ orbitals are fully occupied, while the 2a$^{\prime\prime}$ orbital is empty. The defect can be ionized and we calculated the respective charge transition levels. 

\subsection{Substitutional sulfur}
\label{sec:sub}

By substituting carbon with sulfur, two extra electrons are provided into the lattice.
Our calculation revealed that the substitutional sulfur, S$_\text{s}$, can exist in doubly positive, singly positive, and neutral charge states depending on the position of the Fermi-level.  

In the neutral charge state, the two extra electrons occupy a triple degenerate $t_2$ level in the bandgap. This creates a Jahn-Teller unstable system. We find that the $C_{2v}$ distortion represents the most stable structure, followed by a $C_{3v}$ distorted structure with spin-singlet ground state. This result is corroborated by recent molecular dynamics calculations~\cite{Yu2019}. We also note that similar Jahn-Teller behavior was observed for substitutional nickel defect in diamond~\cite{Thiering2024}.

The electronic structure of the neutral S$_\text{s}$ in the ground state is depicted in Fig.~\ref{fig:ks-states}. This defect, which has a spin-singlet ground state, exhibits a fully occupied a$_1$ and an empty b$_1$ level. Both the fully symmetric a$_1$ and the asymmetric b$_1$ orbitals are localized on the $p$ orbitals of sulfur, with additional contributions from the nearest carbon atoms.

\begin{figure}[!htp]
    \includegraphics[scale=.03]{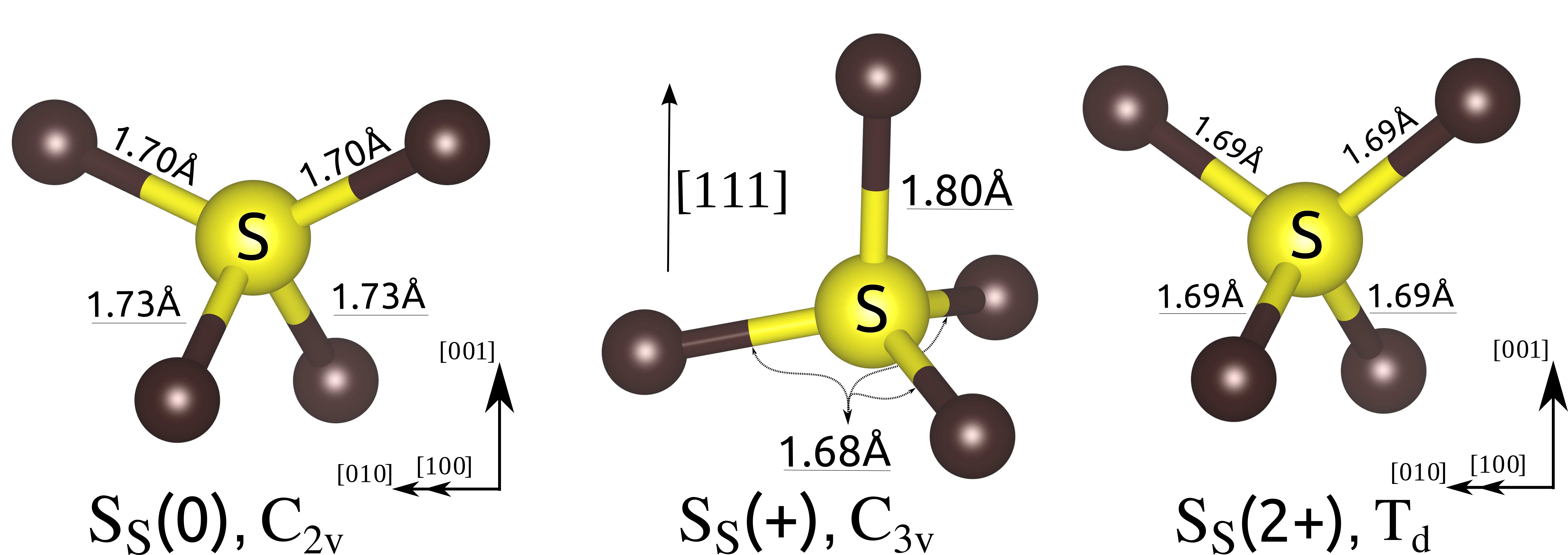}
    \caption{\label{fig:sub_structure} Geometry of optimized sulfur substitutional defect in various charge states. The ($2+$) state is isovalent with carbon resulting in T$_d$ symmetry. Additional electrons distort the geometry because of Jahn-Teller distortion.
 S and C atoms are defined as yellow and dark spheres, respectively.}
\end{figure}

In the single positive charge state, a single electron occupies the triple degenerate $t_2$ level, which again leads to a Jahn-Teller unstable configuration. We find that the $C_{3v}$ geometry is the most stable configuration with $S=1/2$ ground state. However, in the ($2+$) charge state, the defect remains in tetrahedral symmetry (see Fig.~\ref{fig:sub_structure}) with spin-singlet ground state. 

Previous calculations from Wang \textit{et al.}~\cite{Wang2002} reported the ($+|0$) donor level at $E_\text{c} - 0.79$~eV. However, our calculations predict this charge transition level at $E_\text{c} - 1.6$~eV which agrees well with the result of a recent work~\cite{Yu2019}. The calculated second donor level, ($2+|+$), is located at $E_\text{c} - 2.9$~eV.

The S$_\text{s}$($+$) defect is paramagnetic, which can be detected by electron paramagnetic resonance (EPR) spectroscopy. Indeed, a sulfur-related $S=1/2$ EPR center, labeled as W31 center~\cite{Wyk1986}, was observed with trigonal symmetry in diamond. The chemical identification of W31 was based on the relative intensity of the hyperfine lines corresponding to the 0.7\% natural abundance of $^{33}$S, which has a  nuclear spin $I=3/2$. 

\begin{figure}[t]
     \includegraphics[scale=1.1]{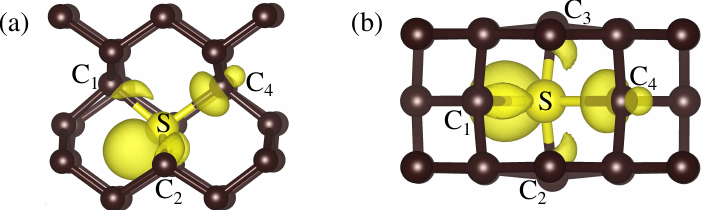} 
     \caption{\label{fig:sub_spindensity} Spin density distribution around the core of the S$_\text{s}$($+$) defect (a) side and (b) top view of
($1\bar{1}0$) plane. S and C atoms are shown by yellow and dark spheres,
respectively. The hyperfine constants for the atoms labeled here are given in Table~\ref{tab:hyper-sv-w31}. Yellow lobes represent a positive isovalues of the
calculated spin density with isosurface value at 0.008~{\AA}$^{-3}$.}
 \end{figure}

Given that S$_\text{s}$($+$) has trigonal symmetry, it could have been associated with the W31 EPR center. The calculated spin density of S$_\text{s}$($+$) defect is depicted in Fig.~\ref{fig:sub_spindensity}. The spin density is distributed around the carbon atoms neighboring the sulfur dopant, where $^{13}$C isotope (with  nuclear spin $I=1/2$ and 1.1\% natural abundance) may also appear as hyperfine satellite lines in the EPR spectrum. Due to the $C_{3v}$ symmetry, the spin density and the hyperfine constants on the three carbon atoms (C$_1$, C$_2$, and C$_3$ in Fig.~\ref{fig:sub_spindensity}) are identical. The observed and computed hyperfine constants are provided in Table~\ref{tab:hyper-sv-w31}. It becomes apparent that the calculated hyperfine constants for $^{33}$S are much smaller than the values observed for the W31 center, which rules out the S$_\text{s}$($+$) defect as a likely candidate for this center. Note that these results are consistent with a previous DFT study~\cite{Baker2008}.

\subsection{Sulfur-vacancy defect}
\label{sec:SV}

During the implantation process, a large number of vacancies are created in the diamond lattice. Typically, thermal annealing is applied to remedy the crystalline structure of diamond, which induces mobility of the vacancies at certain temperatures. A single or multiple mobile vacancies may combine with the substitutional sulfur and form sulfur-vacancy complexes. The simplest of these complexes involves the pairing of a single vacancy with sulfur, which we label as SV defect.

\begin{figure}[h]
    \includegraphics[scale=.07]{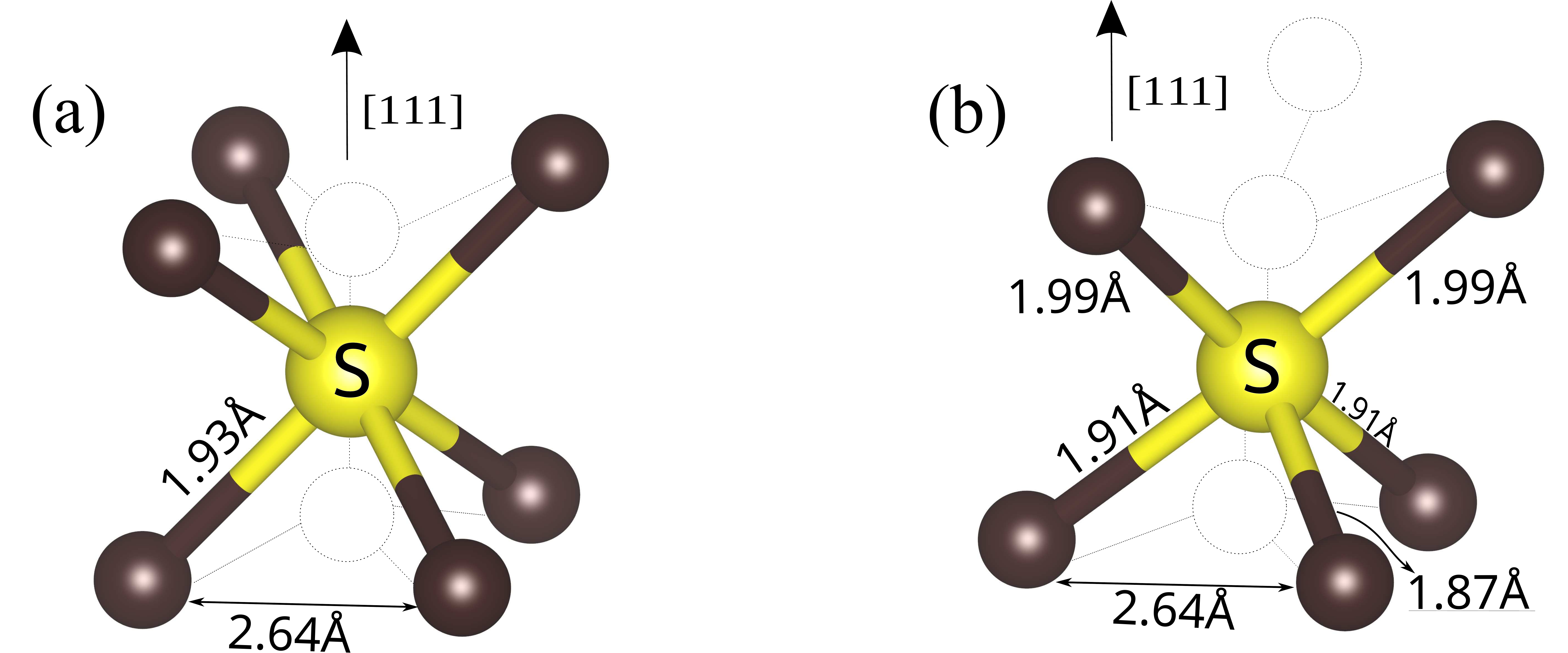}
    \caption{\label{fig:SV-SVV_structure} 
    The optimized geometry of sulfur-vacancy complexes. (a) SV defect and (b) SVV defect. S and C atoms are illustrated as yellow and dark spheres, respectively. The dashed spheres represent the vacancies.}
\end{figure}

Our calculations show that sulfur adjacent to a vacancy 
relaxes to the interstitial site near the vacancy, at the inversion point of the diamond lattice (aka the "split-interstitial" position), where it adopts a $D_{3d}$ symmetry.
In turn, the on-site SV configuration, which exhibits $C_{3v}$ symmetry, lies about 2.64~eV higher in energy than the most stable $D_{3d}$ configuration [see Fig.~\ref{fig:SV-SVV_structure}(a)]. 

In the SV defect, six carbon dangling bonds are created, with sulfur providing six electrons to the lattice. The six dangling bonds give rise to the a$_{2u}$, e$_u$, e$_g$ and a$_{1g}$ defect levels, where the e$_u$ and e$_g$ are double degenerate, whereas sulfur also creates a deep a$_{1g}$ level. In total, twelve electrons occupy the a$_{1g}$, a$_{2u}$, e$_u$ and $e_g$ levels, whereas the upper a$_{1g}$ level remains empty in the neutral charge state. Among these, the fully occupied e$_g$ level appears just above the valence band maximum, whereas the empty a$_{1g}$ level lies in the upper half of the band gap. Due to the parity selection rule, optical transitions between the e$_g$ and a$_{1g}$ are forbidden, making this defect optically inactive.

Furthermore, upon the charge state transition, the empty a$_{1g}$ level may be filled with either one or two electrons. We find that the SV($-$) defect is stable when the Fermi-level lies between $E_\text{v}+4.0$~eV and $E_\text{v}+4.5$~eV (see Fig.~\ref{fig:formation}). If sulfur donors remain in the diamond lattice after sulfur implantation they may stabilize SV($-$) with donating an electron towards the SV defect. 

\begin{figure}[b]
     \includegraphics[scale=1.2]{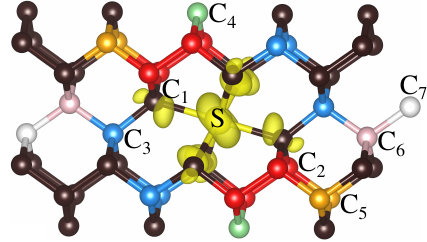} 
     \caption{\label{fig:sv_spindensity} Spin density distribution around the core of the SV($-$) defect with side view respect to (1$\bar{1}$0) plane. S and C atoms are shown by yellow and dark spheres,
respectively. For the second, third, fourth, fifth, and sixth nearest neighbors, we changed the carbon atoms color to red, blue, green, pink, orange, and white, respectively.  The hyperfine constants for the atoms labeled here are given in Table~\ref{tab:hyper-sv-w31}. Yellow lobes represent positive isovalues of the calculated spin density with isosurface value at 0.007~{\AA}$^{-3}$.}
\end{figure}

The SV($-$) is a paramagnetic defect with a ground state of spin $S=1/2$. To characterize its magnetic properties, we calculated both the spin density and hyperfine tensors. The spin density is shown in Fig.~\ref{fig:sv_spindensity}. To distinguish between atoms in different neighbor shells, we assigned unique colors to the atoms from the second to the seventh nearest-neighbor shells. Atoms sharing the same color are equivalent, having identical hyperfine constants. As shown in Fig.~\ref{fig:sv_spindensity}, the first through seventh nearest neighbor shells contain 6, 12, 6, 6, 12, 6, and 6 equivalent carbon atoms, respectively, consistent with the $D_{3d}$ symmetry of the defect. The respective hyperfine constants for these atoms are provided in Table~\ref{tab:hyper-sv-w31}, along with their multiplicities ($n$).

Superhyperfine satellites are observed alongside the  main hyperfine splitting from the $^{33}$S in the W31 EPR center~\cite{Wyk1986}. These satellites can be attributed to hyperfine coupling with $^{13}$C nuclear spins. The study identified four $^{13}$C atoms in the first shell, six in the second shell, and twelve in the third shell, all symmetrically equivalent within their respective shells. In a later theoretical study~\cite{Baker2008}, the W31 EPR center was associated with the SV($-$) defect, with the authors concluding that the number of equivalent $^{13}$C atoms in the first shell is six.  

\begin{figure}[h]
     \includegraphics[scale=.34]{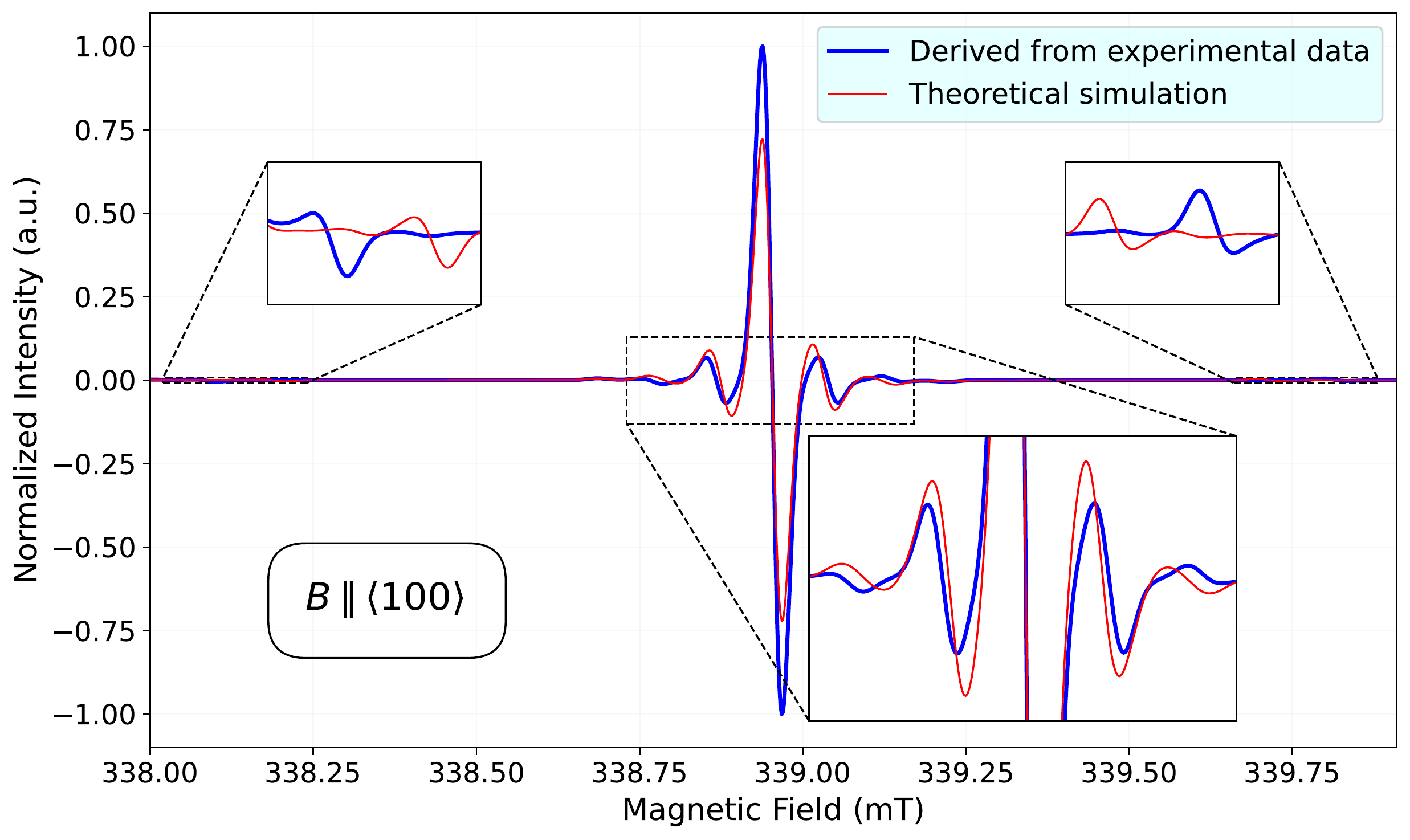} 
     \caption{\label{fig:EPR} Simulated EPR spectrum for SV($-$) defect (red curve) and the EPR spectrum as derived from the spin Hamiltonian of the W31 EPR center (blue curve) from Ref.~\onlinecite{Wyk1986}. The hyperfine splitting from $^{33}$S is not shown here. The respective $^{13}$C hyperfine tensors are listed in Table~\ref{tab:hyper-sv-w31}.}
 \end{figure}

To simulate the EPR spectrum in the X-band, we used the calculated hyperfine tensors, focusing specifically on the $^{13}$C superhyperfine satellites (depicted by the red curve in Fig.~\ref{fig:EPR}). To this end, we employed the \textsc{easyspin} code ~\cite{Stoll2006}. Unfortunately, the direct comparison with the observed EPR spectrum is extremely difficult because  the quality of the plot, reported in Ref.~\onlinecite{Wyk1986} is barely reproducible and we were unable to find any other publications with showing the spectrum of the W31 EPR center. Therefore, we rather used the spin Hamiltonian, which was fitted to the experimental data as given in Ref.~\onlinecite{Wyk1986} (e.g., hyperfine tensors and multiplicity of $^{13}$C spins), to generate a derived EPR spectrum (blue curve in Fig.~\ref{fig:EPR}).

\begin{table}[h]
\caption{\label{tab:hyper-sv-w31}
Hyperfine tensors for the W31 EPR center (MHz) (experimental data are taken from Ref.~\onlinecite{Wyk1986}) and the SV($-$) and S$_\text{s}$($+$) defects calculated by HSE06 functional in diamond. The atom labels are shown in Fig.~\ref{fig:sv_spindensity}.}
\begin{ruledtabular}
\begin{tabular}{llllr}
 Atom & $n$ & $A_{\parallel}$  & $A_{\perp}$   & $A_{\parallel}$ ($\theta, \phi) $   \\
\hline 
W31 & $S=1/2$ & & & \\
S       & 1 & 1029.0 & 1033.7 & 54.7, 45 \\
C$_1$ & 4 & 70.6   & 45.1   & 61, $-$135   \\
C$_2$ & 6 & 14.9 & 9.8 & $<$111$>$ \\
C$_3$ & 12 & 4.8 & 4.8 & Isotropic \\
\arrayrulecolor{gray}\hline
SV($-$) & $S=1/2$    &  & \\
S       & 1  & 1120.7 & 1135.8 & 54.7, 45       \\
C$_{1}$ & 6  & 69.4   & 42.2   & 59.2, $-$135   \\
C$_{2}$ & 12 & $-$4.5 & $-$3.5 & 54.1, $-$42.2  \\
C$_{3}$ & 6  & 15.2   & 8.6    & 47.3, 45       \\
C$_{4}$ & 6  & 7      & 5.5    & 174.7, $-$135.2\\
C$_{5}$ & 12 & 4.3    & 3.0    &  55.1, 49.7    \\
C$_{6}$ & 6  & 11.4   & 8.1    &  53.6,$-$135.0 \\
C$_{7}$ & 6  & 5.7    & 4.2    &  59.3, 45      \\
\arrayrulecolor{gray}\hline
S$_\text{s}$($+$) & $S=1/2$    &  & \\
S       & 1  &  136.9  & 27.5   & 54.7, 45   \\
C$_{(1-3)}$ & 3  &  70.7  &  52.2  & 36, $-$135   \\
C$_{4}$ & 1  &  218.3  &  167  & 54.7, 45   \\
\end{tabular}
\end{ruledtabular}
\end{table}

Based on the good agreement between the experimental and simulated spectra, we infer that the W31 EPR center is associated with the SV($-$) defect which agrees with the result of a previous study~\cite{Baker2008}. To further support this assignment, the EPR spectrum of the W31 center should be revisited by carefully recording the spectrum in the region of the main electron spin transition, in order to accurately obtain the respective relative intensity between the main EPR line and the hyperfine satellites. 

In the experimental study, the W31 EPR center was found in Type 1b diamond~\cite{Wyk1986} which contains a relatively high concentration of nitrogen impurities. However, the mechanism by which sulfur was incorporated into the diamond in that study remains unclear~\cite{Wyk1986}. Both substitutional nitrogen (N$_\text{s}$) and sulfur may act as donors (see Fig.~\ref{fig:formation}) that can provide an electron towards the SV defect to form SV($-$). If S$_\text{s}$ defects coexist with isolated vacancies in the sample, annealing may initiate the formation of SV complex when mobile vacancies are trapped by the S$_\text{s}$.  

\begin{figure*}[t]
 \includegraphics[scale=0.9]{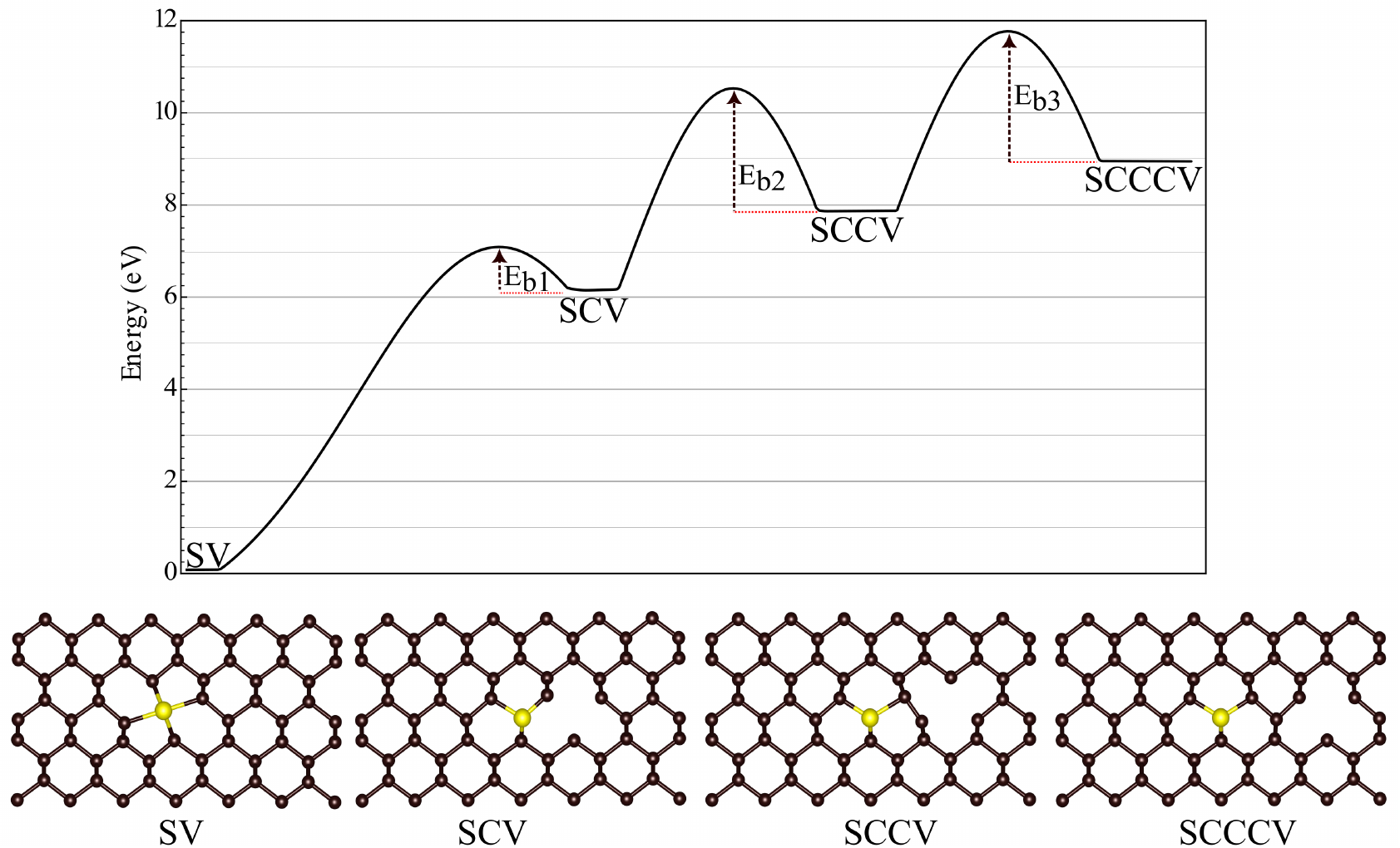}
    \caption{\label{fig:barrier} 
    Geometry of the defects after optimization along [110] mirror plane for SV, SCV, SCCV, and  SCCCV. S and C atoms are yellow and dark spheres, respectively. The diffusion barrier energy are $E_\text{b1}=0.87$~eV, $E_\text{b2}= 2.70$~eV, and $E_\text{b3}= 2.77$~eV. }
\end{figure*}

We further proceed by studying the trapping mechanism of a vacancy by the S$_\text{s}$ through \textit{ab initio} calculations. To this end, the vacancy was placed at the fourth neighbor shell along the $(110)$ plane (labeled as SCCCV), and we used $\langle 100 \rangle$ direction as a configuration coordinate to map the adiabatic potential energy surface as the vacancy approaches the S$_\text{s}$. The resulting configurations were labeled as SCCV and SCV, and the final configuration was the SV defect as depicted in Fig.~\ref{fig:barrier}.

Our calculations reveal that the diffusion barrier energy  reduces successively as the vacancy approaches the S$_\text{s}$. The calculated highest diffusion barrier energy almost coincides with the value for the diffusion of the neutral vacancies in diamond (2.8~eV) using the HSE06 functional~\cite{Deak2014}. These results imply that the mobile vacancy is effectively trapped by the S$_\text{s}$ and the SV defect can be readily formed. The main driving forces behind the formation of the complex are the attractive strain field towards the vacancies, generated by the S$_\text{s}$, as well as the charge transfer between the species where the electron flows from the S$_\text{s}$ towards the vacancy. 

The large energy gain ($\approx 8$~eV) in the formation of the complex can be understood as the elimination of the high-energy dangling bonds of the vacancy and the reduction in local strain around the SV defect relative to the S$_\text{s}$. In the most stable split-vacancy configuration of the SV defect, the six dangling bonds interact strongly with the sulfur ion, as we already discussed in our analysis of the electronic structure of the SV defect. 

We note that we performed analogous calculations for the complex formation between oxygen and vacancy in a previous study~\cite{Ghafari2023}. In that work, we found a metastable OCCV configuration~\cite{Ghafari2023}, using the same labeling scheme for the oxygen complex as we did for the sulfur complex. A similar metastable structure was also found for complexes of two vacancies, VCCV~\cite{Slepetz2014}. However, the sulfur-vacancy complexes do not follow this pattern. Specifically, the SCCV has a higher formation energy than the SCV defect. We attribute this difference to the large strain field generated by the S$_\text{s}$, which stabilizes the SCV relative to the SCCV. In contrast, no large strain field is present for the O$_\text{s}$ or the vacancy defect, making their configurations behave differently.

\subsection{Sulfur-vacancy-vacancy defect}
\label{sec:SVV}

In sulfur-implanted samples, multiple vacancies are formed, leading to the potential trapping of another vacancy by the SV defect, thereby creating the SVV complex. Therefore, we considered the SVV defect in our study where the optimized structure of the neutral defect is depicted in Fig.~\ref{fig:SV-SVV_structure}(b).

Four defect levels appear in the band gap due to the SVV (see Fig.~\ref{fig:ks-states}). The a$^{\prime}$ and a$^{\prime \prime}$ occupied levels are mainly localized on the carbon dangling bonds, while the empty ones are the carbon-sulfur antibonding orbitals mixed with the carbon-centered orbitals at the second vacancy site farther from the sulfur ion. As the a$^{\prime}$ and a$^{\prime \prime}$ levels are close in energy, a high-spin state $S=1$ is stabilized in the neutral charge state. The occupied and empty defect levels in the gap indicate that SVV may occur in various charge states. Indeed, we find that ($+$), ($0$), ($-$), and ($2-$) charge states can exist, while the ($-$) state of $S=1/2$ spans a wide range of the Fermi energies in the upper half of the bandgap (see Fig.~\ref{fig:formation}). 

\begin{figure}[h]
     \includegraphics[scale=1.1]{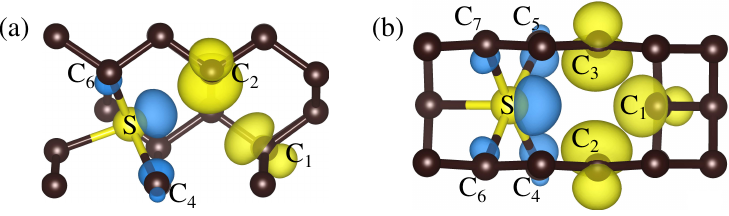}
     \caption{\label{fig:SVV_spindensity} Spin density distribution around the core of the SVV($-$) defect with two different views: (a) side and (b) top view of
(1$\bar{1}$0)  plane. S and C atoms are shown by yellow and dark spheres,
respectively. The hyperfine constants for the atoms labeled here are given in Table~\ref{tab:hyper-sv-w31}. Yellow lobes represent positive isovalues while blue lobes represent negative isovalues of the calculated spin density with an isosurface value at 0.01~{\AA}$^{-3}$.}
\end{figure}

In both the neutral and negative charge states, the SVV defect is paramagnetic, with the total spin values of $S = 1$ and $S = 1/2$, respectively.
The calculated spin density in the negative charge state is shown in Fig.~\ref{fig:SVV_spindensity}, where we labeled the atoms for which the hyperfine tensors are provided in Table~\ref{tab:hyper-svv}. In the neutral charge state, the spin density is mostly distributed over the C$_{1-3}$ atoms. Due to the C$_{1h}$ symmetry, the spin density (and consequently hyperfine constants) on the C$_2$ and C$_3$ atoms (similarly on C$_4$ and C$_5$ and on C$_6$ and C$_7$ for the negative charge state) are identical.
In the neutral charge state, a marginal spin density is observed on the sulfur atom; instead, the spin is predominantly localized on the sp$^3$ hybridized dangling bonds of the C$_{1-3}$ carbon atoms. In contrast, upon the introduction of an additional electron to form the SVV($-$), a redistribution of spin density occurs, which results in significant spin localization on the sulfur atom and its four adjacent carbon atoms.

\begin{table}[h]
\caption{\label{tab:hyper-svv}
Hyperfine constants ($A_{xx}$, $A_{yy}$ and $A_{zz}$) of the sulfur atom and the nearest neighbor atoms for the neutral and negatively charged SVV defects. The atom labels are shown in Fig.~\ref{fig:SVV_spindensity}.}
\begin{ruledtabular}
\begin{tabular}{lllr}
 Atom &  $A_{xx}$ (MHz) & $A_{yy}$ (MHz)  &  $A_{zz}$ (MHz)  \\
\hline
SVV($0$)  & \multicolumn{3}{l}{$S=1/2$} \\
S         & 50.1  & 49.3 & 56.3 \\
C$_1$     & 86.6  & 86.1 & 173.7 \\ 
C$_{2-3}$ & 85.2  & 84.6 & 159.4 \\
SVV($-$)  & \multicolumn{3}{l}{$S=1$} \\
S         & $-374.1$  & $-372.4$ & $-419.5$ \\
C$_1$     & 162.8  & 161.8 & 264.7 \\ 
C$_{2-3}$ & 194.7  & 194.3 & 354.4 \\
C$_{4-5}$ & $−50.9$  & $−49.2$ & $−83.4$ \\
C$_{6-7}$ & $−33.1$  & $−29.4$ & $−56.9$ \\
\end{tabular}
\end{ruledtabular}
\end{table}


\subsection{Complexes of hydrogen and substitutional sulfur}

To proceed, we considered the complexes of hydrogen with sulfur defects which may occur in CVD diamonds. First, we present the results on the complex of hydrogen and substitutional sulfur, i.e., the SH defect. SH defects were previously studied by semilocal DFT calculations, and it was concluded that the most stable SH defect acts as a shallow donor~\cite{Lombardi2004}. 

\begin{figure}[h]
    \includegraphics[scale=1.0]{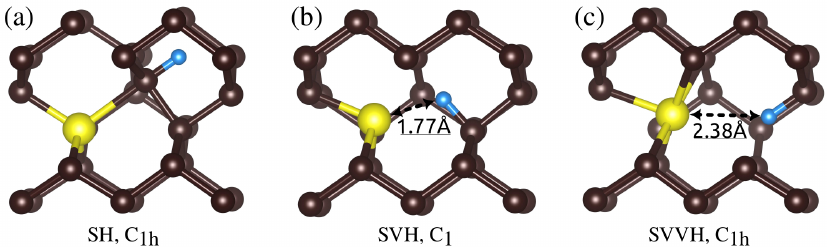}
    \caption{\label{fig:SVH_structure} The optimized geometry of (a) SH, (b) SVH, (c) SVVH complexes. S, H, and C atoms are illustrated as yellow, blue, and dark spheres, respectively.}
\end{figure}

\begin{figure}[h]
     \includegraphics[scale=1.1]{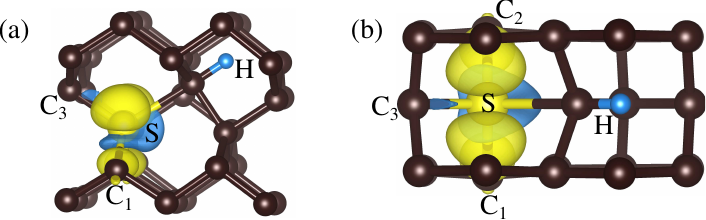} 
     \caption{\label{fig:sh_spindensity} Spin density distribution around the core of the SH defect with two different views: (a) side  and (b) top view of
(1$\bar{1}$0)  plane. S and C atoms are shown by yellow and dark spheres,
respectively. The hyperfine constants for the atoms labeled here are
given in Table~\ref{tab:hyper-sh}. Yellow lobes represent a positive isovalues and blue lobes present a negative isovalues in the calculated spin density with an isosurface value at 0.009~{\AA}$^{-3}$.}
\end{figure}

With substitutional sulfur, the hydrogen interstitial atom (H$_\text{i}$) can be positioned either at the antibonding or bond-centered site. Our calculations show that the antibonding site is the most stable configuration for the SH defect [see Fig.~\ref{fig:SVH_structure}(a)], while the bond-centered position is $0.91$~eV higher in energy. This finding is consistent with the results of previous calculations~\cite{Nishimatsu2001, Lombardi2004}.
This defect with an almost C$_{1h}$ symmetry and a doublet ground state $S=1/2$ generates three levels in the bandgap (see Fig.~\ref{fig:ks-states}), with the spin density depicted in Fig.~\ref{fig:sh_spindensity}. 
We further calculated the hyperfine tensors for the $^{1}$H, $^{13}$C, and $^{33}$S nuclei with neural abundance 99.98\%, 1.1\%, and 0.75\%, respectively.  Because of the C$_{1h}$ symmetry, the spin density and the hyperfine constants on the C$_1$ and C$_2$ atoms are identical. The hyperfine constants are presented in Table~\ref{tab:hyper-sh}.
\begin{table}[h]
\caption{\label{tab:hyper-sh}
Hyperfine constants ($A_{xx}$, $A_{yy}$, and $A_{zz}$) for the nearest-neighbor atoms for the neutral SH defect. The atom labels are shown in Fig.~\ref{fig:sh_spindensity}.}
\begin{ruledtabular}
\begin{tabular}{lllr}
 Atom &  $A_{xx}$ (MHz) & $A_{yy}$ (MHz)  &  $A_{zz}$ (MHz)  \\
\hline 
S         & $-27.5$  & $-20.0$ & $57.3$  \\
H         & $-11.4$  & $-7.1$  & $-13.9$ \\ 
C$_{1-2}$ & $150.9$  & $149.8$ & $183.9$ \\ 
C$_{3}$   & $-7.0$   & $-5.7$  & $-8.9$  \\
\end{tabular}
\end{ruledtabular}
\end{table}

Previous semilocal DFT calculations predicted~\cite{Lombardi2004} that the SH defect has a relatively shallow donor level at $E_\text{c} - 0.61$~eV. However, semilocal DFT methods underestimate the fundamental bandgap of diamond by about 1~eV which makes this prediction ambiguous. Indeed, our HSE06 calculations result in the donor level at $E_\text{c} - 1.2$~eV which is relatively deep, though it is a shallower donor than the S$_\text{s}$ (see Fig.~\ref{fig:formation}).


\subsection{Complexes of hydrogen and sulfur-vacancy defects}
\label{sec:S-V-H defects}
\subsubsection{Sulfur-Vacancy-Hydrogen complex}

The diffusing hydrogen interstitial may also be trapped by the SV defect, which is the most stable sulfur-related defect in diamond. By adding one hydrogen to the SV defect, one of the sulfur-carbon bonds breaks, and the system adopts a $C_{1h}$ symmetry [see Fig.~\ref{fig:SVH_structure}(b)]. As a result, the e$_u$ level splits into a$^\prime$ and a$^{\prime \prime}$ levels (see Fig.~\ref{fig:ks-states}). 
The neutral charge state of the SVH defect is stable, when the Fermi-level is located in the lower half of the band gap, whereas the negative charge state is the most stable for the Fermi-level in the upper half of the band gap.

\begin{figure}[h]
     \includegraphics[scale=1.2]{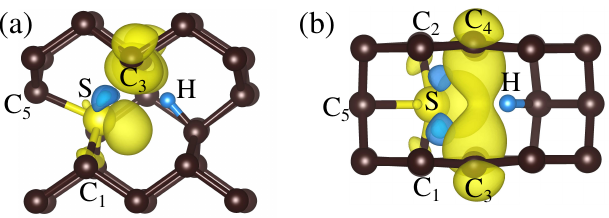} 
     \caption{\label{fig:svh_spindensity} Spin density distribution around the SVH defect with two different views: (a) side  and (b) top view of
(1$\bar{1}$0)  plane. S and C atoms are shown by yellow and dark spheres,
respectively. The hyperfine constants for the atoms labeled here are
given in Table~\ref{tab:hyper-svh}. Yellow and blue lobes represent a positive and negative isovalues of the calculated spin density with an isosurface value at 0.006~{\AA}$^{-3}$, respectively.}
 \end{figure}

The neutral charge state of the SVH defect has an odd number of electrons with a ground state $S=1/2$, with strong spin density localization on the sulfur atom (see Fig.~\ref{fig:svh_spindensity}). As a consequence, the SVH($0$) defect may be observed by electron spin magnetic resonance techniques with characteristic hyperfine satellites. A large hyperfine splitting is observable for the $^{33}$S, similar to the case of SVV($-$) defect (c.f., Table~\ref{tab:hyper-svv}). However, the EPR spectra of the two defects can be readily distinguished by the hyperfine doublet of $^1$H; the hyperfine coupling is small, because the spin density has a tiny contribution from the hydrogen atom. Unlike the case of SVV($-$), all the hyperfine couplings of $^{13}$C ions are much smaller than those of $^{33}$S, which also enables to distinguish the spectra of the two defects.

\begin{table}[h]
\caption{\label{tab:hyper-svh}
Hyperfine constants ($A_{xx}$, $A_{yy}$ and $A_{zz}$) for the nearest-neighbor atoms for the neutral SVH defect. The atom labels are shown in Fig.~\ref{fig:svh_spindensity}.}
\begin{ruledtabular}
\begin{tabular}{lllr}
 Atom &  $A_{xx}$ (MHz) & $A_{yy}$ (MHz)  &  $A_{zz}$ (MHz)  \\
\hline 
S         & 368.9  & 363.3 & 429.8 \\
H         &   6.1  & 0.5   & 12.1  \\ 
C$_{1}$   &  12.0  & 11.4  & 24.4  \\ 
C$_{2}$   &  14.0  & 13.5  & 27.8  \\ 
C$_{3}$   &  83.0  & 81.5  & 168.3 \\
C$_{4}$   &  76.8  & 75.4  & 155.0 \\
C$_{5}$   & $-10.35$ &  $−8.9$ & $-13.3$ \\
\end{tabular}
\end{ruledtabular}
\end{table}

\subsubsection{Sulfur-Vacancy-Vacancy-Hydrogen complex}

The SVV defect possesses multiple dangling bonds and one of them may be passivated by a mobile interstitial hydrogen in diamond. The resulting SVVH defect, shown in Fig.~\ref{fig:SVH_structure}(c), has a $C_{1h}$ symmetry. Due to multiple occupied and empty defect levels in the bandgap (Fig.~\ref{fig:ks-states}), various charge states of SVVH occur as a function of the Fermi-level (Fig.~\ref{fig:formation}). The neutral charge state is dominant with the ground state $S=1/2$ for the position of the Fermi-level in the lower half of the bandgap, whereas the negative charge state is dominant with a ground state $S=1$ for the position of the Fermi-level in the upper half of the band gap. The high-spin ground state is stabilized in a similar fashion as it occurred for the neutral SVV defect.

\begin{figure}[h]
     \includegraphics[scale=1.2]{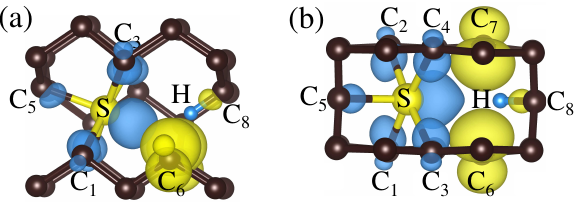} 
     \caption{\label{fig:svvh_spindensity} Spin density distribution around the core of the neutral SVVH defect with two different views: (a) side  and (b) top view of
(1$\bar{1}$0)  plane. S and C atoms are shown by yellow and dark spheres,
respectively. The hyperfine constants for the atoms labeled here are given in Table~\ref{tab:hyper-svvh}. Yellow and blue lobes represent a positive and negative isovalues at 0.0076~{\AA}$^{-3}$ of the calculated spin density, respectively. }
\end{figure}

The spin density for the neutral SVVH defect is shown in Fig.~\ref{fig:svvh_spindensity}. 
Because of the C$_{1h}$ symmetry, the spin density and the hyperfine constants on the two sides of our mirror plane are identical. A blue lobes indicate a negative spin density. The majority of the spin density is localized on the carbon dangling bonds of the C$_{6-7}$ atoms. The hyperfine constants are listed in Table~\ref{tab:hyper-svvh}. The spin density is significant on the C-H unit of the defect, yielding the absolute values of the hyperfine constants of $^1$H close to or above $10$~MHz. In the negative charge state, the spin density is mostly localized on the S-atom; the largest hyperfine constants can be observed for the $^{33}$S, which are, however, much smaller than those for the SV($-$), SVV($-$) or SVH($0$) defects (c.f.\ Tables~\ref{tab:hyper-sv-w31}-\ref{tab:hyper-svvh}). The second largest hyperfine constants occur for two identical $^{13}$C atoms. The hyperfine coupling of $^1$H is weak, yet observable, with the absolute values below $10$~MHz. 

\begin{table}[h]
\caption{\label{tab:hyper-svvh}
Hyperfine constants ($A_{xx}$, $A_{yy}$ and $A_{zz}$) for the nearest-neighbor atoms for the neutral and negatively charged SVVH defects. The atom labels are shown in Fig.~\ref{fig:svvh_spindensity}.}
\begin{ruledtabular}
\begin{tabular}{lllr}
 Atom &  $A_{xx}$ (MHz) & $A_{yy}$ (MHz)  &  $A_{zz}$ (MHz)  \\
\hline 
SVVH($0$) & \multicolumn{3}{l}{$S=1/2$} \\
S         & $−106.4$ & $−105.3$ & $-138.7$  \\
H         &  17.8  &  8.87  & $-28.49$  \\ 
C$_{1-2}$ & $−29.2$  & $−25.4$  & $−48.2$  \\ 
C$_{3-4}$ & $−23.3$  & $−22.1$  & $−40.9$   \\ 
C$_{5}$   & $−24.9$  & $−24.5$  & $-35.2$   \\
C$_{6-7}$ &  131.8 &  131.0 & 307.1   \\
C$_{8}$   &  17.6  &  16.3  & 34.3    \\
SVVH($-$) & \multicolumn{3}{l}{$S=1$} \\
S         &  252.4  & 248.5 &  274.5 \\
H         & $-6.2$    & 5.8   & 8.5 \\ 
C$_{1-2}$ & 12.3 & 10.7 & 20.6 \\
C$_{3-4}$ & 36.4 & 35.8 & 59.8 \\
C$_{5}$   & $-9.0$ & $-8.8$ & $-12.2$ \\
C$_{6-7}$ & 80.5 & 80.4 & 166.3 \\
C$_{8}$   & 3.4  &  3.1 & 9.6  \\
\end{tabular}
\end{ruledtabular}
\end{table}


\subsection{Description of sulfur implanted diamonds}
\label{sec:sulfur_implantation}

Before considering the implications of implantation, we discuss the thermal stability of the basic sulfur defects, by comparing their formation energies, shown 
in the left panel of Fig.~\ref{fig:formation}. Except for the p-type diamond, the SV complex evolves as the most stable defect, followed by the S$_\text{s}$ defect. The interstitial defect has a very high formation energy and it is unlikely to occur in diamond. Indeed, the W31 EPR center, which we have associated with SV($-$) defect, was observed in nitrogen-doped diamond, where the nitrogen-related deep donors may provide electrons to the SV defect. To our knowledge, the W31 EPR center is the only confirmed spectroscopical fingerprint of sulfur in diamond. 

In sulfur-implanted diamond, sulfur may create several vacancies in a cascade process, and the sulfur atom may stop at the substitutional site by kicking out a carbon atom from the lattice, forming the S$_\text{s}$ defect. After annealing, the vacancies may diffuse depending on the annealing temperature. Neutral vacancies, V($0$), become mobile at temperatures around 600$^{\circ}$C, with an activation energy ($\Delta E$) of $2.3\pm 0.3$~eV, as determined experimentally~\cite{Davies1992}, which was also largely confirmed by the HSE06 calculations~\cite{Deak2014}. For the V($-$), the HSE06 diffusion activation energy of 3.5~eV was obtained~\cite{Deak2014}. The diffusion constant can be computed as follows:
\begin{equation}
\label{eq:activation}
    \mathcal{D} \approx \exp (-\frac{\Delta E}{k_\text{B} T}) \text{,}
\end{equation}
where $k_\text{B}$ is the Boltzmann constant and $T$ is the temperature.
By assuming the same diffusion constant for V($0$) and V($-$) defects in Eq.~\eqref{eq:activation}, it can be estimated that V($-$) defects become mobile from temperatures around 800$^{\circ}$C. 

To achieve mobile neutral vacancies, the (quasi) Fermi-level is to be positioned at around $E_\text{v} + 2.0$~eV (see Table~\ref{tab:chargetransitionlevel}) that may appear in heavily implanted regions of diamond. At that Fermi-level, the S$_\text{s}$ is in the ($2+$) charge state and the resulting SV defect becomes neutral. Since multiple vacancies are formed during implantation, another vacancy may approach the neutral SV defect forming the neutral SVV defect. The binding energy of these defect reactions can be computed as
\begin{align} 
     \text{V}(0) + \text{S}_\text{s} (2+) & \Rightarrow  \text{SV}(0) + 2h + 8.7  \: \text{eV} \label{eq:sv} \\
     \text{V}(0) + \text{SV}(0)  & \Rightarrow  \text{SVV}(0) + 2.5  \: \text{eV,} \label{eq:svv} 
\end{align}
where $h$ denotes the hole, generated in the process. Eq.~\eqref{eq:sv} implies that the formation of SV defect is very likely when the vacancy is near S$_\text{s}$. On the other hand, the binding energy in Eq.~\eqref{eq:svv} between SV($0$) and V($0$) is close to the barrier energy of V($0$), thus SVV($0$) may dissociate upon annealing. 

We also note that two neutral vacancies may be combined to form divacancies,
\begin{align}
     \text{V}(0) + \text{V}(0) &\Rightarrow  \text{V}_2(0) + 4.2  \: \text{eV.}  \label{eq:vac}
\end{align}
These divacancies may survive the annealing process and could appear in the sulfur implanted diamonds as residual defects. On the other hand, the results imply that sulfur is preferred over the partner vacancy for the diffusing vacancies as the energy gain towards SV is much larger than that towards V$_2$. Therefore, if the S$_\text{s}$ is situated near multiple vacancies in a region of sulfur implanted diamond then SV preferably forms over V$_2$.

\begin{figure*}[ht]
    \includegraphics[scale=0.87]{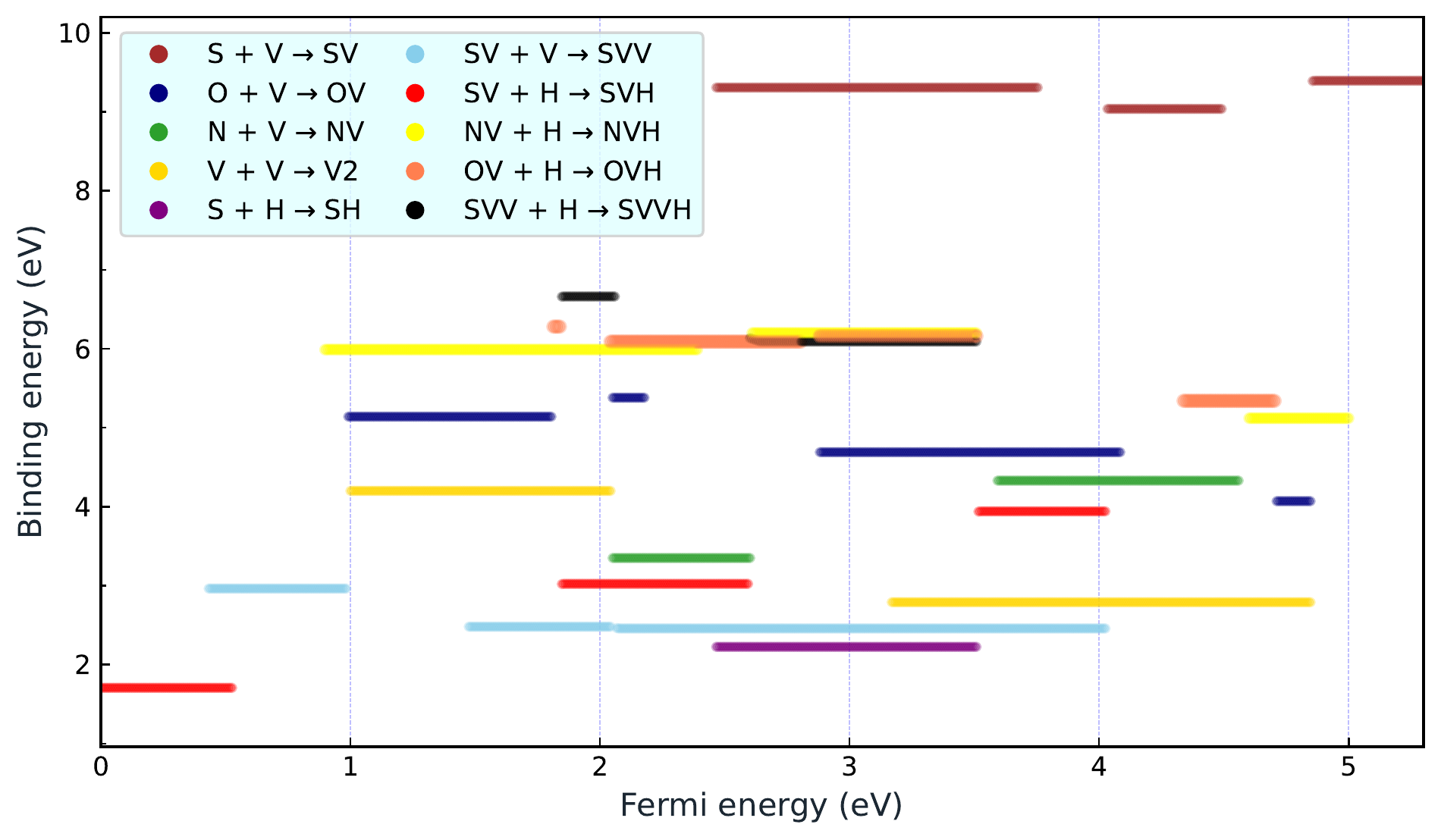}
    \caption{\label{fig:binding} The binding energy for complex formation of SV, OV, NV, V$_2$,  SVV, SVV, SVH, NVH, and OVH as a function of the Fermi-level. We here consider charge conserving defect reactions without ejecting holes or electrons. Furthermore, constituent defect species with sharing the same sign of charge states are disregarded due to Coulomb repulsion.}
\end{figure*}

In the experiment~\cite{Luhmann2018}, the sulfur implantation is carried out to CVD diamond samples with a depth of about 50~nm. In these diamond samples, it is likely that hydrogen is left in the sample, which can be activated and become mobile upon annealing. The most stable sulfur-related SV defect may be combined with interstitial hydrogen, which is in turn neutral at $E_\text{Fermi} = E_\text{v} + 2.0$~eV (see right panel of Fig.~\ref{fig:formation} and Table~\ref{tab:chargetransitionlevel}). As shown in Fig.~\ref{fig:binding}, the calculated binding energy for this process is 3.0~eV. Furthermore, the defect reaction between the S$_\text{s}$ and the H$_\text{i}$ reads as
\begin{align}
     \text{S}_\text{s}(2+) + \text{H}_\text{i} & \Rightarrow  \text{SH}(+) + h + 1.7  \: \text{eV.} \label{eq:sh}
\end{align}
This indicates that the SH defect is unstable upon annealing at the temperatures when the neutral vacancies are mobile.

In the same experiment, the annealing temperature was set to 1200~$^\circ$C after sulfur implantation~\cite{Luhmann2018} where the vacancies are definitely mobile. Our results imply that the majority of sulfur defects appear as the SV while the S$_\text{s}$ also occurs but to a smaller extent. Furthermore, the SV may adsorb interstitial hydrogen, so that the SVH defect could also appear in CVD diamonds.
Beside those defects, divacancies (or larger vacancy aggregates) also exist where few isolated vacancies could appear ejected from larger vacancy aggregates before the annealing has been stopped. The schema of sulfur implanted diamonds is depicted in the right panel of Fig.~\ref{fig:schematic} as a summary.

\begin{table}[hbt!]
\caption{\label{tab:chargetransitionlevel}
The charge transition levels of defects. Donor levels are given with respect to $E_\text{c}$ (conduction band minimum), and acceptor levels are given with respect to $E_\text{v}$ (valence band maximum). The data related to oxygen defects are taken from Ref.~\onlinecite{Thiering2016} whereas the data related to intrinsic and nitrogen defects are taken from Ref.~\onlinecite{Deak2014}.}
\begin{ruledtabular}
\begin{tabular}{llr}
 Defect & charge transition level & Energy  (eV)   \\
\hline 
S$_\text{i}$ & ($2+|+$)  & $E_\text{c}- 3.8$  \\
             & ($+|0$)   & $E_\text{c}- 2.6$  \\ 

S$_\text{s}$ & ($2+|+$) &  $E_\text{c}- 2.9$  \\
             & ($+|0$)  &  $E_\text{c}- 1.6$  \\

SV & ($0|-$)  &  $E_\text{v}+4.0$   \\
   & ($-|2-$) &  $E_\text{v}+4.5$   \\

SVV &  ($+|0$)  &  $E_\text{c}- 4.1$ \\
    &  ($0|-$)  &  $E_\text{v}+ 2.3$ \\
    &  ($-|2-$) &  $E_\text{v}+ 3.9$ \\

SH &   ($+|0$)  &  $E_\text{c}- 1.2$ \\

SVH &   ($+|0$)  &  $E_\text{c}- 4.8$ \\
    &   ($0|-$)  &  $E_\text{v}+ 2.6$ \\

SVVH &   ($+|0$)  &  $E_\text{c}- 4.8$ \\
    &   ($0|-$)  &  $E_\text{v}+ 2.6$ \\
    &   ($-|2-$) &  $E_\text{v}+ 4.3$ \\

H$_\text{i}$ &   ($+|0$)  &  $E_\text{c}- 3.5$ \\
             &   ($0|-$)  &  $E_\text{v}+ 3.5$ \\

O$_\text{s}$ & ($2+|+$) &  $E_\text{c}- 4.9$ \\
             & ($+|0$)  &  $E_\text{c}- 3.2$ \\
             & ($0|-$)  &  $E_\text{v}+ 4.1$ \\
             & ($-|2-$) &  $E_\text{v}+ 5.0$ \\

OV & ($2+|+$) &  $E_\text{c} - 4.9$  \\
   & ($+|0$)  &  $E_\text{c} - 3.6$  \\
   & ($0|-$)  &  $E_\text{v} + 2.9$  \\
   & ($-|2-$) &  $E_\text{v} + 4.7$  \\
   
OVH &  ($+|0$)  &  $E_\text{c}- 3.3$ \\
             & ($0|-$)  &  $E_\text{v}+2.8$ \\
             & ($-|2-$) &  $E_\text{v}+ 4.3$ \\

N$_\text{s}$ & ($+|0$) &  $E_\text{c}- 1.8$ \\
             & ($0|-$) &  $E_\text{c}- 3.6$ \\

NV  & ($+|0$)  &  $E_\text{c} - 4.4$   \\
    & ($0|-$)  &  $E_\text{v} + 2.7$   \\
    & ($-|2-$) &  $E_\text{v} + 4.9$   \\

NVH  & ($+|0$)  &  $E_\text{c} - 3.5$   \\
    & ($0|-$)  &  $E_\text{v} + 2.4$   \\
    & ($-|2-$) &  $E_\text{v} + 2.9$   \\

V  & ($2+|+$) &  $E_\text{c} - 4.9$   \\
   & ($+|0$)  &  $E_\text{c} - 4.4$   \\
   & ($0|-$)  &  $E_\text{v} + 2.0$   \\
   & ($-|2-$) &  $E_\text{v} + 4.9$   \\

V$_\text{2}$ & ($+|0$)  & $E_\text{c} - 4.3$  \\
             & ($0|-$)  & $E_\text{v} + 2.3$  \\
             & ($-|2-$) & $E_\text{v} + 3.2$  \\   
\end{tabular}
\end{ruledtabular}
\end{table}


\subsection{NV creation in sulfur-doped diamond with nitrogen implantation}
\label{sec:nv-sulfur}

Nitrogen implantation can be used to create NV centers as single photon emitters in sulfur doped diamonds. In this case, the concentration of nitrogen is much lower than that of the dopants. During nitrogen implantation, a significant number of vacancies are created. If these vacancies remain neutral, they tend to aggregate and form larger vacancy complexes such as V$_2$ defects, which may lower the efficiency of NV creation. 

As shown in previous studies, the charge states of defects in diamond depend on the presence of nearby donors or acceptors~\cite{Collins2002, Manson2018}. If nitrogen is implanted relatively close to the S$_\text{s}$, the latter may induce a negative charge state on two vacancies generated by nitrogen implantation. Note that the S$_\text{s}$ cannot donate both electrons to the same vacancy since the S$_\text{s}$(2+) and the V($2-$) do not share the same Fermi-level region. We note that it has been demonstrated that phosphorus doping induces a negative charge on vacancies~\cite{Luhmann2019}, where phosphorus is a single-electron donor. In this term, the S$_\text{s}$ is more efficient as a double donor.

During annealing the mobile negatively charged vacancies do not recombine with each other but either with the N$_\text{s}$ or the S$_\text{s}$. In the first case, negatively charged NV defect, i.e., the NV center is formed. In the second case, SV($-$) defect is formed. It is difficult to estimate the ratio of the two processes. The binding energy for the complex formation of NV defect is read as
\begin{align} 
     \text{V}(0) + \text{N}_\text{s}(+) & \Rightarrow  \text{NV}(0) + h + 4.2  \: \text{eV,} \label{eq:nv}
\end{align}
which is significantly smaller than that for the SV formation. Nevertheless, if the vacancies are created closer to the N$_\text{s}$ than the S$_\text{s}$ after nitrogen implantation, then the N$_\text{s}$ may trap the diffusing vacancy before it can combine with the S$_\text{s}$. 

In total, sulfur doped CVD diamond mediates the formation of NV center after nitrogen implantation by the following factors: (i) SV defects trap interstitial hydrogen, therefore, many interstitial hydrogen defects are removed prior or after nitrogen implantation and, as a consequence, the formation of NVH defects~\cite{Glover2003} (see Fig.~\ref{fig:formation}) becomes less likely with keeping the NV centers intact; (ii) S$_\text{s}$ defects are double donors that can effectively charge vacancies formed after nitrogen implantation which prevents the formation of vacancy clusters by increasing the probability to combine diffusing vacancies with N$_\text{s}$. 


It is intriguing to analyze the photostability and spin coherence of NV center in S-doped CVD diamonds. The NV center measurements are carried out under the illumination of green light (typically, achieved with a 2.33-eV laser). To this end, the photoionization threshold energies should be studied for the most important sulfur-related defects. When the defect looses an electron in the photoionization process, the threshold energy is referenced to the conduction band minimum, whereas it is referenced to the valence band maximum for adding an electron to the defect in the photoionization process.  

We start this analysis with the S$_\text{s}$ defect. If the defect is initially in the neutral charge state, it will rapidly transition to the ($+$) charge state upon green illumination. In this charge state, the green light does not induce any charge flipping (see Table~\ref{tab:chargetransitionlevel}). Similarly, if the defect is in the ($2+$) charge state then the green light has no effect on its charge state. As discussed above, the most likely stable charge state of the S$_\text{s}$ is ($2+$), which is diamagnetic. Therefore, no spin or charge fluctuations occur due to S$_\text{s}$ defects.

Next, we turn to the neutral SV defect, which is optically inactive because of the selection rules, making it a photostable defect. Furthermore, if the defect is in the negative charge state, the 2.33-eV excitation converts it to neutral charge state. In either case, the SV defect remains in the non-magnetic neutral charge state. Therefore, the SV defect provides an ideal environment for quantum optics operation of NV centers in diamond.

Additionally, we consider the SVH defect under illumination. The neutral charge state remains stable, while the negative charge state will convert to the neutral state when excited by the 2.33-eV laser. If the neutral SVH defect captures a hole, the positive charge state will remain photostable. Hence, we conclude that the most likely configuration for the SVH defect is the neutral charge state, which has a $S=1/2$ paramagnetic state.  
\begin{figure*}[ht]
\includegraphics[scale=.64]{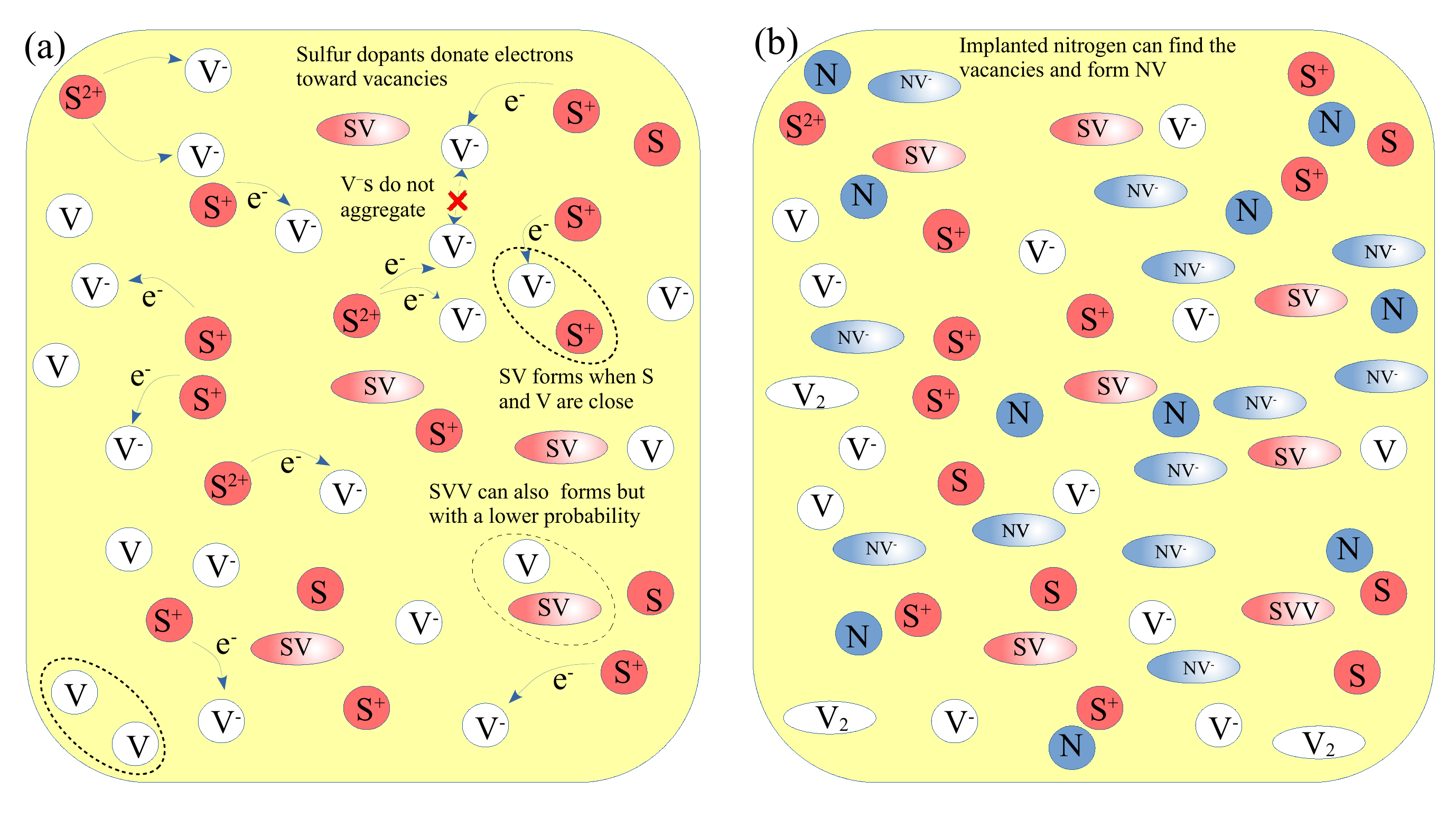} 
\caption{\label{fig:schematic} Schematic diagram about sulfur-doped diamond via sulfur implantation and annealing (a) before and (b) after nitrogen implantation with annealing.}
\end{figure*}

The schematic diagram in Fig.~\ref{fig:schematic}(a) illustrates the possible processes in sulfur-doped diamond.  A critical issue with sulfur doping is that the SV complex is principally more stable than the S$_\text{s}$, limiting the ability to fully realize the double donor nature of S$_\text{s}$. As a result, not all the implanted sulfur ions can contribute to the activation of NV centers, which could explain the observed 75\% creation yield of NV center after nitrogen implantation of sulfur-doped diamond~\cite{Luhmann2018}, as shown in Fig.~\ref{fig:schematic}(b).

Experimental results have shown that sulfur doping improves the coherence time of NV center's electron spin with respect to that of non-doped CVD diamond~\cite{Luhmann2018}. A reduction in coherence time is typically associated with electron spins in the lattice, generated by defects such as the V$_2$. On the other hand, SV defects have a closed shell singlet ground state that is optically inactive, thus it provides an electron spin-free environment for the NV center in diamond.

\subsection{NV creation in oxygen-doped diamonds with nitrogen implantation}
\label{sec:oxygen}

In this last part, we compare the oxygen-doped and sulfur-doped CVD diamonds in terms of activation of NV centers after nitrogen implantation and annealing, given the observed NV creation yield of 68\% in the oxygen-doped diamond samples~\cite{Luhmann2018}. To this end, we use the computational results on oxygen-related defects from our previous study~\cite{Thiering2016}. The formation energies and charge transition levels of oxygen-related defects are presented in the right panel of Fig.~\ref{fig:formation}, whereas the binding energies of oxygen-related defects are shown in Fig.~\ref{fig:binding}.  

In contrast to sulfur dopant, oxygen, as a light element, is most stable as a substitutional defect (O$_\text{s}$). Although the O$_\text{s}$ has two deep donor levels, it has hyperdeep acceptor levels too, making it an amphoteric defect. The donor levels of O$_\text{s}$ lie deeper than those of S$_\text{s}$. Thus, the O$_\text{s}$ does not negatively charge nearby vacancies as efficiently as the S$_\text{s}$ does. After oxygen ion implantation and annealing, a mobile vacancy may combine with the O$_\text{s}$, forming the OV defect (see Fig.~\ref{fig:binding}). We assume that the OV defect may coexist with the O$_\text{s}$ but the relative concentration between O$_\text{s}$ and OV is much higher that that between S$_\text{s}$ and SV.

The OV complex is also an amphoteric defect, possessing multiple donor and acceptor levels (see Table~\ref{tab:chargetransitionlevel}), where the oxygen remains at the substitutional site, exhibiting the $C_{3v}$ symmetry. The most relevant neutral and negative charge states realize the $S=1$ and $S=1/2$ ground states, respectively~\cite{Thiering2016}. Trapping the mobile interstitial hydrogen further leads to the formation OVH complex, which binding energy is similar to that of NVH complex (see Fig.~\ref{fig:binding}). The OVH complex has both donor and acceptor levels, as well, where the neutral and negative charge states possess the $S=1/2$ and $S=1$ ground states, respectively~\cite{Thiering2016}. 

We hypothesize that the beneficial effect of oxygen-doping for the creation of NV center by nitrogen implantation technique is related to the following factors: 

(i) The majority of implanted oxygen ions substitutes carbon in the lattice, which can negatively charge nearby vacancies. As compared to sulfur-doped diamond, the number of O$_\text{s}$ donors per implanted oxygen ions is higher than the respective value for the S$_\text{s}$. However, the doping efficiency toward vacancies is rather favored by S$_\text{s}$ donors. Hence, the interplay of these two effects ultimately sets the probability of charging nearby vacancies, which is a crucial process in the activation of NV centers in diamond. 

(ii) A fractional part of the oxygen defects  appears as the OV complex. 
In CVD diamonds, OV complexes capture hydrogen interstitials, which decreases the probability of the formation of NVH complexes and leaves NV centers intact. 

Finally, we discuss the photostability of the  oxygen-related defects under the green laser.
The O$_\text{s}$ appears as neutral or single positively charged defect where the latter occurs when charging of a nearby vacancy was successful. 
The ($+$) charge state of O$_\text{s}$ can be converted to the neutral state with ejecting a hole, which might be captured by the NV($-$) defect. Then NV($0$) will be converted to the NV($-$) with ejecting a hole upon strong green illumination. Since the O$_\text{s}$ is neutral in this moment the capture of hole is not so likely, thus the O$_\text{s}$ remains neutral with $S=0$ state. That is a photostable state because green illumination does not convert it to either ($+$) or ($-$) charge state. 

The OV complex may exist in the neutral or ($-$) charge state after defect reactions. Both charge states are photostable upon green illumination. 
Similarly, the OVH complex may exist in the neutral or ($-$) charge state after defect reactions. Both charge states are again photostable. 

These results imply that oxygen doping most likely creates a photostable and electron spin-free environment toward the NV center. The most abundant O$_\text{s}$ defects remains in the $S=0$ ground state, whereas the less abundant oxygen vacancy-related defects are paramagnetic but also photostable. We conclude that the coherence time of the NV centers can be improved in this environment, which supports the earlier experimental findings~\cite{Luhmann2018}.

\section{Conclusions}
\label{sec:conclusion} 

In this paper, we have studied the sulfur defects in diamond by utilizing hybrid density functional theory plane wave supercell calculations. Our investigation focused on various sulfur-related defects, including interstitial sulfur, substitutional sulfur, and sulfur substitutional with vacancy as a complex defect as well as the complexes of these defects with hydrogen. We established the accurate donor and acceptor levels of these defects in diamond. We found that substitutional sulfur is not a shallow donor in diamond. Its complex with hydrogen produces a somewhat shallower donor level but it is still deep. Furthermore, the binding energy of the complex indicates that the complex can dissociate at such annealing temperatures that are typically used to remove vacancy defects in irradiated diamond. Our calculations support previous findings, which associated the W31 EPR center with the negatively charged sulfur-vacancy defect in diamond. We also found that sulfur-vacancy complex can effectively trap mobile interstitial hydrogen in diamond, forming a sulfur-vacancy-hydrogen complex.

We analyzed the possible defect species in sulfur ion implanted and annealed diamond. We concluded that the formation of sulfur-vacancy complex is more likely than that of substitutional sulfur and the sulfur-vacancy complex may trap hydrogen in shallow CVD diamond layers. Implanting nitrogen ions into these sulfur-doped diamonds can facilitate the formation of the nitrogen-vacancy center by the double donor nature of the substitutional sulfur and trapping of hydrogen by the sulfur-vacancy complex. We find that sulfur defects are photostable and provide electron spin-free environment toward the nitrogen-vacancy center, which can principally improve the coherence time of the nitrogen-vacancy center's electron spin. We carried out the same analysis for the oxygen-doped CVD diamond layers by taking our previous \textit{ab initio} data. We find some subtle differences in the mediation of nitrogen-vacancy defect creation by the two dopants. Direct estimation of the creation yield of the nitrogen-vacancy center in sulfur or oxygen-doped diamond is not feasible from the dataset we have, and it is beyond the scope of this study. Nevertheless, our calculations provide atomistic insights about the doped layers and could support the improvement in the creation yield and the electron spin coherence times in these doped diamond layers. Our results imply that the creation yield of nitrogen-vacancy center in sulfur-doped diamond might be improved by careful engineering of the annealing parameters after sulfur implantation, in order to avoid complex formation of substitutional sulfur and the vacancy. This is extremely challenging because substitutional sulfur is less stable than sulfur-vacancy complex but it might be possible to freeze the sulfur defects in their metastable state and simultaneously heal the crystalline damages created by irradiation.

\begin{acknowledgments}
Discussions with S\'ebastien Pezzagna are appreciated. Support by the Ministry of Culture and Innovation and the National Research, Development and Innovation Office within the Quantum Information National Laboratory of Hungary (Grant No.\ 2022-2.1.1-NL-2022-00004) as well as the EU QuantERA II MAESTRO project (NKFIH grant no.\ 2019-2.1.7-ERA-NET-2022-00045), and the European Commission for the projects QuMicro (Grant No.\ 101046911) and SPINUS (Grant No.\ 101135699) are much appreciated. AG acknowledges the high-performance computational resources provided by KIF\"U (Governmental Agency for IT Development of Hungary). A.P. acknowledges the financial support of János Bolyai Research Fellowship of the Hungarian Academy of Sciences

\end{acknowledgments}

\bibliography{bibliography}
\end{document}